\documentclass[useAMS,usenatbib]{mn2e}
\usepackage{graphicx}

\def\be{\begin{equation}} 
\def\ee{\end{equation}}

\def\HI{\hbox{H~$\scriptstyle\rm I\ $}} 
\def\HII{\hbox{H~$\scriptstyle\rm II\ $}} 
\def\HeI{\hbox{He~$\scriptstyle\rm I\ $}} 
\def\HeII{\hbox{He~$\scriptstyle\rm II\ $}} 
\def\HeIII{\hbox{He~$\scriptstyle\rm III\ $}}

\def\gsim{\lower.5ex\hbox{\gtsima}} 
\def\lsim{\lower.5ex\hbox{\ltsima}} \def\gtsima{$\; \buildrel > \over 
\sim \;$} \def\ltsima{$\; \buildrel < \over \sim \;$} \def\prosima{$\; 
\buildrel \propto \over \sim \;$} \def\gsim{\lower.5ex\hbox{\gtsima}} 
\def\lsim{\lower.5ex\hbox{\ltsima}} 
\def\simgt{\lower.5ex\hbox{\gtsima}} 
\def\simlt{\lower.5ex\hbox{\ltsima}} 
\def\simpr{\lower.5ex\hbox{\prosima}} \def\la{\lsim} \def\ga{\gsim} 
  
 \def\gtsima{$\; \buildrel > \over \sim \;$} 
\def\ltsima{$\; \buildrel < \over \sim \;$} 
\def\gsim{\lower.5ex\hbox{\gtsima}} 
\def\lsim{\lower.5ex\hbox{\ltsima}} 
\def\simgt{\lower.5ex\hbox{\gtsima}} 
\def\simlt{\lower.5ex\hbox{\ltsima}} 
\def\simpr{\lower.5ex\hbox{\prosima}} 
\def\la{\lsim} 
\def\ga{\gsim}

\def\Lya{Ly$\alpha$~}

\def\E3{{\cal E}_{\rm g}^{III}}

\def\Msun{\rm M_\odot}

\def\fesc{f_{\rm esc}}
\def\Lya{Ly$\alpha$ }
\def\avchi{$\langle \chi_{HI} \rangle$ }
\def\lobs{erg s$^{-1}$}
\title[LAE visibility]{The visibility of Lyman Alpha Emitters: constraining reionization, ionizing photon escape fractions and dust} 
\author[Hutter et al.]{Anne Hutter$^{1}$\thanks{E-mail:ahutter@aip.de}, Pratika Dayal$^{2}$, Adrian M. Partl$^{1}$, Volker M\"uller$^{1}$ \\ 
$^{{1}}$ Leibniz-Institut f\"ur Astrophysik, An der Sternwarte 16, 14482 Potsdam, Germany\\
$^{2}$ SUPA\thanks{Scottish Universities Physics Alliance}, Institute for Astronomy, University of Edinburgh, Royal Observatory, Edinburgh, EH9 3HJ, UK}

\begin{document} 
 
\date{} 
 
%\pagerange{\pageref{firstpage}--\pageref{lastpage}} \pubyear{2009} 
 
\maketitle 
 
\label{firstpage} 
\begin{abstract} 
We build a physical model for high-redshift Lyman Alpha emitters (LAEs) by coupling state of the art cosmological simulations (GADGET-2) with a dust model and a radiative transfer code (pCRASH). We post-process the cosmological simulation with pCRASH using five different values of the escape fraction of hydrogen ionizing photons ($\fesc=0.05,0.25,0.5,0.75,0.95$) until reionization is complete, i.e. the average neutral hydrogen fraction drops to \avchi$\simeq 10^{-4}$. 
Then, the only free-parameter left to match model results to the observed \Lya and UV luminosity functions of LAEs at $z\simeq6.6$ is the relative escape of Lyman Alpha (\Lya) and continuum photons from the galactic environment ($f_\alpha/f_c$). We find a {\it three-dimensional} degeneracy such that the theoretical model can be reconciled with observations for an IGM \Lya transmission $\langle T_{\alpha} \rangle_{LAE}\simeq 38-50$\% (which translates to \avchi$\simeq 0.5-10^{-4}$ for Gaussian emission lines), $\fesc \simeq 0.05-0.50$ and $f_\alpha/f_c\simeq0.6-1.8$. 

\end{abstract}


% ***************************************************************************************************************
\section{Introduction}
\label{intro}
% ***************************************************************************************************************
The Epoch of Reionization (EoR) begins when the first stars start producing neutral hydrogen (\HI) ionizing photons, carving out an ionized Str\"omgren region in the neutral intergalactic medium (IGM) around themselves, and ends when all the \HI in the IGM is ionized. Understanding the process of cosmic reionization is extremely important because in addition to marking the last major change in the ionization state of the Universe, it affected all subsequent structure formation through a number of radiative feedback effects \citep[see e.g.][and references therein]{barkana-loeb2001, ciardi-ferrara2005, maio2011, sobacchi2013, wyithe2013}. A broad picture that has emerged \citep{choudhury-ferrara2007} is one wherein hydrogen reionization is an extended process that starts at $z \approx 15$ and is about 90\% complete by $z \simeq 8$; while it is initially driven by metal-free Population III stars in low-mass halos ($\leq 10^8\Msun$), the conditions for star formation in these halos are soon erased by a combination of chemical and radiative feedback by $z \simeq 10$. However, any attempt at modelling reionization necessarily requires a number of important assumptions regarding the number of \HI ionizing photons produced by the source galaxy population depending on their physical properties, the fraction of these photons ($\fesc$) that can escape out of the galactic environment and contribute to reionization, and the clumping factor of the IGM at high-redshifts, to name a few \citep{salvaterra2011}. Given these assumptions, reionization scenarios need to be constantly updated as new data sets are acquired. 

By virtue of their continually growing numbers, a valuable new data set is provided by high-redshift Lyman Alpha Emitters (LAEs); this is a class of galaxies identified by means of their Lyman Alpha (Ly$\alpha$) emission line at 1216\,\AA\, in the galaxy rest-frame. Indeed, hundreds of LAEs have now been confirmed in the epoch of reionization: $z \simeq 5.7$ \citep{malhotra2005, shimasaku2006, hu2010, curtislake2012}, $z \simeq 6.6$ \citep{taniguchi2005, kashikawa2006, hu2010, ouchi2010, kashikawa2011} and $z \simeq 7$ \citep{iye2006, ouchi2009, stark2010,pentericci2011}. In addition to their number statistics, LAEs have been rapidly gaining popularity as probes of reionization and high-redshift galaxy evolution for two reasons: (a) the strength, width and continuum break bluewards of the \Lya line makes their detection unambiguous, and (b) \Lya photons are extremely sensitive to attenuation by \HI; the observed \Lya luminosity can then be used to infer the ionization state of the IGM at redshifts close to those of the emitter and hence to reconstruct, at least piecewise, the cosmic reionization history. 

However, interpreting LAE data is complicated by a number of physical effects. First, the intrinsic \Lya luminosity depends on the total number of \HI ionizing photons that are produced by a galaxy, depending on the star formation rate (SFR), age and metallicity of its stellar population \citep[e.g.][]{santos2004}. Second, depending on the \HI and dust contents in the interstellar medium (ISM), only a fraction ($1-\fesc$) of these \HI ionizing photons are able to contribute to the intrinsic \Lya luminosity by ionizing the ISM \HI; the rest contribute to building the ionized \HII region around the galaxy. We briefly digress to note that the dependence of $\fesc$ on galaxy properties has been the subject of much recent debate: while some authors find $\fesc$ to decrease with an increase in the halo mass \citep{razoumov2010,yajima2011,ferrara2013}, other works find the opposite trend \citep{gnedin2008,wise2009}. The value of $\fesc$ also remains poorly constrained with findings ranging from a few percent  \citep[e.g.][]{gnedin2008} up to $20-30$\% \citep[e.g.][]{mitra2013} or even higher \citep[e.g.][]{wise2009}. Third, only a fraction ($f_\alpha$) of the \Lya photons produced inside a galaxy can escape out of it unattenuated by dust \citep{dayal2008, finkelstein2009, nagamine2010, forero2011, dayal2012}. Fourth, only a fraction ($T_\alpha$) of the \Lya photons that emerge out of a galaxy are transmitted through the IGM and reach the observer. As expected, this transmission sensitively depends on the IGM \HI ionization state and it has been shown that only galaxies residing in over-lapping \HII regions would be observed as LAEs in the initial stages of reionization, i.e. reionization increases the observed clustering of LAEs \citep{mcquinn2007b, dayal2009}. Fifth, the \Lya IGM transmission calculation is complicated by the presence of peculiar velocities. Inflows (outflows) of gas into (from) the emitter blue-shift (red-shift) the \Lya line, leading to a decrease (increase) in the value of $T_\alpha$ along different lines of sight, strongly affecting the visibility of galaxies as LAEs \citep{verhamme2006, iliev2008, zheng2010, dijkstra2011}. Indeed, \citet{santos2004} has shown that the IGM ionization state can not be constrained in the presence of peculiar velocities. However, \citet{dayal2011} have shown that the interpretation of LAE data is much more involved; a decrease in the \Lya transmission due to peculiar velocities and/or a highly neutral IGM can be compensated by an increase in $f_\alpha$ due to dust being clumped in the ISM of LAEs, as per the `Neufeld-model' \citep{neufeld1991}.

A number of past theoretical works have used one or more of the above ingredients to use LAEs as tracers of reionization. With semi-analytic modelling, a number of authors \citep[e.g.][]{dijkstra2007, dayal2008, samui2009} have shown that the LAE \Lya luminosity functions (LFs) are consistent with a fully ionized IGM and can be explained solely by an evolution of the underlying dark matter halo mass functions. However, when considered in combination with the non-evolving observed ultraviolet (UV) LFs of LAEs between $z \simeq 5.7$ and $6.6$ \citep{kashikawa2006}, some of these works \citep{dijkstra2007,ouchi2010} argue for an additional dimming of the Ly$\alpha$ line by about $30$\%.
A number of studies have used large pure dark matter simulations to study the clustering of LAEs \citep{mcquinn2007, iliev2008, orsi2008}. Finally, a number of studies have been undertaken using cosmological hydrodynamic simulations with dust \citep{dayal2008,dayal2009,nagamine2010,forero2011}, with radiative transfer without dust \citep{zheng2010} and radiative transfer with dust \citep{dayal2011,forero2011,duval2014}. 

This work is quite close in spirit to the calculations presented in \citet{dayal2011} wherein the authors used (a) cosmological hydrodynamic simulations run with GADGET-2 to obtain the physical properties of $z \simeq 5.7$ galaxies, (b) a dust model that took into account the entire star formation history of each galaxy to calculate its dust enrichment and, (c) a RT code (CRASH) to obtain the ionization fields to calculate the IGM \Lya transmission for each galaxy, in order to identify the simulated galaxies that would be observationally classified as LAEs. Using this model, the authors showed that the effects of dust and IGM are degenerate in affecting the visibility of LAEs such that a large $f_\alpha$ can be compensated by a small $T_\alpha$ and vice versa to yield a given value of the observed \Lya luminosity. 
However, the main caveat in that work was that the authors used a constant value of $\fesc=0.02$ \citep{gnedin2008} for all galaxies in their calculations, and started their RT runs assuming the IGM gas to be in photoionization equilibrium with a uniform ultraviolet background (UVB) produced by unresolved sources corresponding to an average neutral hydrogen fraction, $\langle \chi_{\rm HI} \rangle= 0.3$. 

In this work, we substiantially enhance the model presented in \citet{dayal2011} to build a self-consistent model that couples cosmological SPH simulations, dust modelling and a fast RT code (pCRASH) to identify LAEs without making any prior assumptions on the IGM ionization state, $\fesc$ and the ISM dust distribution. Starting from a uniform neutral IGM, consistent with the \citet{haardt-madau1996} background imposed in simulations, we use 5 different values of $\fesc=0.05,0.25,0.5,0.75,0.95$ in order to study how varying $\fesc$ affects the visibility of LAEs through: (a) the direct impact on the intrinsic \Lya luminosity, and (b) the IGM \Lya transmission that depends on the topology and extent of \HII regions as determined by $\fesc$. Comparing the statistics of the simulated LAEs to observations, our aim is to jointly constrain three fundamental parameters that determine the visibility of LAEs - $\fesc$, $\chi_{HI}$ (or $T_\alpha$) and the relative escape of \Lya photons with respect to continuum photons from the dusty ISM of galaxies ($f_\alpha/f_c$).

We begin by describing the hydrodynamical simulation, the dust model and the identification of galaxies as Lyman Break Galaxies (LBG) through the UV continuum in Sec. \ref{sec2}. We follow this approach since UV photons are only attenuated by dust but unaffected by the IGM ionization state, simplifying their identification. In order to validate the simulations, we compare the simulated LBG UV LFs to observations in Sec. \ref{sec3}; a comparison of the simulated LBG stellar mass functions, stellar mass densities and specific star formation rates with the observations are shown in Appendix \ref{a2} for completeness. Once the physical properties of the simulated galaxies have been validated, we choose a simulation snapshot at the edge of the epoch of reonization ($z \simeq 6.7$) and post-process it with a fast RT code (pCRASH) with the 5 values of $\fesc$ mentioned above, as explained in Sec. \ref{sec4}. We then compare the simulated LAE UV and \Lya LFs with observations \citep{kashikawa2011} in Sec. \ref{sec5} in order to jointly constrain $\fesc$, $\chi_{HI}$ and $f_\alpha/f_c$.
Due to the model dependent relation between $\chi_{HI}$ and the \Lya IGM transmission ($T_{\alpha}$) we show our constraints also in terms of $\fesc$, $T_{\alpha}$ and $f_\alpha/f_c$, before concluding in Sec. \ref{sec6}.

% ***************************************************************************************************
\section{Cosmological hydrodynamic simulations}
\label{sec2}
% ***************************************************************************************************
In this section we describe the cosmological simulation used to obtain the physical properties of high-redshift galaxies and the semi-analytic model used to obtain their dust enrichment, in order to calculate their observed visibility in the UV.

% ***************************************************************************************************
\subsection{The simulation}
\label{hydro_sims}
% ***************************************************************************************************
The hydrodynamical simulation analyzed in this work has been carried out using the TreePM-SPH code GADGET-2 \citep{springel2005}. The adopted cosmological model corresponds to the $\Lambda$CDM universe with dark matter (DM), dark energy and baryonic density parameter values of ($\Omega_{\Lambda}$, $\Omega_m$, $\Omega_b$) = (0.73, 0.27, 0.047), a Hubble constant $ H_0= 100h = 70 {\rm km s^{-1} Mpc^{-1}}$, and a normalisation $\sigma_8=0.82$, consistent with the results from WMAP5 \citep{komatsu2009}. The simulation box has a size of $80h^{-1}$ comoving Mpc (cMpc), and contains $1024^3$ DM particles, and initially the same number of gas particles; the mass of a DM and gas particle is $3.6 \times 10^7 h^{-1} \Msun$ and $6.3 \times 10^6 h^{-1} \Msun$, respectively. The softening length for the gravitational force is taken to be $3 h^{-1}$ comoving kpc and the value of the smoothing length for the SPH kernel for the computation of hydrodynamic forces is allowed to drop at most to the gravitational softening.

The runs include the star formation prescriptions of \citet{springel-hernquist2003b} such that the ISM is described as an ambient hot gas containing cold clouds, which provide the reservoir for star formation, with the two phases being in pressure equilibrium; the relative number of stars of different masses is computed using the Salpeter initial mass function \citep[IMF;][]{salpeter1955} between 0.1-100$\Msun$. The density of the cold and of the hot phase represents an average over small regions of the ISM, within which individual molecular clouds cannot be resolved by simulations sampling cosmological volumes. The runs also include the feedback model described in \citet{springel-hernquist2003b} which includes (a) thermal feedback: supernovae (SN) inject entropy into the ISM, heat up nearby particles and destroy molecules, (b) chemical feedback: metals produced by star formation and SN are carried by winds and pollute the ISM, and (c) mechanical feedback: galactic winds powered by SN. In the case of mechanical feedback, the momentum and energy carried away by the winds are calculated assuming that the galactic winds have a fixed velocity of $500 \, {\rm km \, s^{-1}}$ with a mass upload rate equal to twice the local star formation rate, and carry away a fixed fraction ($50\%$) of the SN energy (for which the canonical value of $10^{51} \,{\rm ergs}$ is adopted). Finally, the run assumes a metallicity-dependent radiative cooling \citep{sutherland-dopita1993} and a uniform redshift-dependent UV Background (UVB) produced by quasars and galaxies as given by \citet{haardt-madau1996}. 

Galaxies are recognized as gravitationally-bound groups of at least 20 total (DM+gas+star) particles using the Amiga Halo Finder \citep[AHF;][]{knollmann2009}. When compared to the standard Sheth-Tormen mass function \citep{sheth-tormen1999}, the simulated mass function is complete for halo masses $M_h \geq 10^{9.2} \, \Msun$ for $z \simeq 6-8$; galaxies above this mass cut-off are referred to as the ``complete sample''. Of this complete sample, we identify all the ``resolved" galaxies that contain a minimum of $4N$ gas particles, where $N=40$ is the number of nearest neighbours used in the SPH computations; this is twice the standard value of $2N$ gas particles needed to obtain reasonable and converging results \citep[e.g.][]{bate-burkert1997}. We impose an additional constraint and only use those resolved galaxies that contain at least 10 star particles so as to get a statistical estimate of the composite spectral energy distribution (SED). For each resolved galaxy used in our calculations (with $M_h \geq 10^{9.2} \, \Msun$, more than $4N$ gas particles and a minimum of 10 star particles) we obtain the properties of all its star particles, including the redshift of, and mass/metallicity at formation; we also compute global properties including the total stellar mass ($M_*$), gas mass ($M_g$), DM mass ($M_h$), mass-weighted stellar metallicity ($Z_*$) and the mass weighted stellar age ($t_*$). 

% ***************************************************************************************************
\subsection{Dust model}
\label{dust}
% ***************************************************************************************************
The evidence for dust at high-redshifts comes from observations of damped Ly$\alpha$ systems \citep{pettini1994, ledoux2002} and from the thermal dust emission from SDSS QSOs \citep{omont2001, bertoldi-cox2002}. Although dust is produced both by SN and evolved stars, several works \citep{todini-ferrara2001, dwek2007} have shown that the contribution of AGBs to the total dust content is negligible at $z \gsim 6$, since the age of the Universe is shorter than the typical evolutionary timescales of AGBs above this redshift. We therefore make the hypothesis that Type II SN (SNII) are the primary dust factories and compute the total dust mass, $M_d$ in each of our simulated galaxies as described in \citet{dayal2010a}.

\begin{figure*}
\begin{center}
\center{\includegraphics[width=1.0\textwidth]{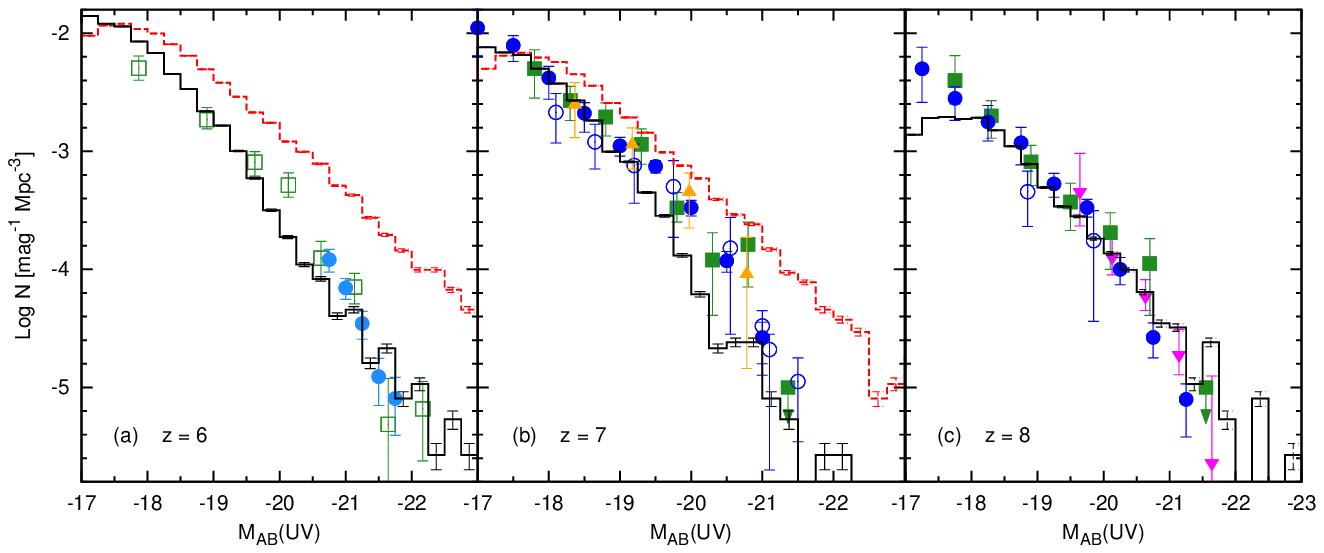}}
  \caption{UV luminosity functions at $z \simeq 6, 7$ and $8$, from left to right, as marked in each panel. In all panels, the solid black (dashed red) histograms show the dust corrected (intrinsic) simulated UV LFs with error bars showing the Poissonians error and symbols showing the observed data. The observed UV LFs have been taken from (a) $z \simeq 6$: \citet[][empty squares]{bouwens2007}, \citet[][filled circles]{mclure2009}; (b) $z \simeq 7$: \citet[][filled squares]{bouwens2011b}, \citet[][empty circles]{mclure2010}, \citet[][filled circles]{mclure2013}, \citet[][filled triangles]{oesch2010} and (c) $z \simeq 8$: \citet[][filled triangles]{bradley2012}, \citet[][filled squares]{bouwens2011b}, \citet[][empty circles]{mclure2010} and \citet[][filled circles]{mclure2013}.}
 \label{fig_uvlf}
\end{center}
\end{figure*}

This dust mass can be used to obtain the total optical depth, $\tau_c$, to continuum photons as
\begin{equation}
\tau_c = \frac{3\Sigma_d}{4 a s},
\end{equation}
where $\Sigma_d = M_d [\pi r_d^2]^{-1}$ is the dust surface mass density, $r_d$ is the dust distribution radius, $a=0.05 \mu m$ and $s = 2.25\, {\rm g\, cm^{-3}}$ are the radius and material density of graphite/carbonaceous grains, respectively \citep{todini-ferrara2001}. Since in our model dust and gas are assumed to be perfectly mixed, the dust distribution radius is taken to be equal to the gas distribution radius $r_g = 4.5\lambda r_{vir}$ where the spin parameter ($\lambda$) has a value of about 0.04 averaged across the galaxy population studied \citep{barnes1987,steinmetz1995, ferrara2000} and $r_{vir}$ is the virial radius, calculated assuming the collapsed region has an overdensity of 200 times the critical density. This dust optical depth can then be used to obtain the escape fraction of UV photons ($f_c$) assuming a screen-like dust distribution such that $f_c = e^{-\tau_c}$. 

% ***************************************************************************************************
\subsection{Identifying Lyman Break Galaxies}
\label{identify_lbgs}
% ***************************************************************************************************
To identify the simulated galaxies that could be observed as LBGs at $z \simeq 6-8$, we start by computing their UV luminosities. We consider each star particle to form in a burst, after which it evolves passively. The total SED, including both the stellar and nebular continuum, is computed for each star particle via the population synthesis code {\tt STARBURST99} \citep{leitherer1999}, using its mass, stellar metallicity and age. The total intrinsic UV luminosity, $L_c^{int}$ (at 1500 \AA\, in the galaxy rest frame) is then calculated for each progenitor by summing the SEDs of all its star particles. 

Continuum photons can be absorbed by dust within the ISM and only a fraction, $f_c$, escape out of any galaxy unattenuated by dust. However, these photons are unaffected by the ionization state of the IGM, and all the continuum photons that escape out of a galaxy can then reach the observer, so that the observed continuum luminosity can be expressed as $L_c^{obs} = L_c^{int} \times f_c$; this can be translated into an absolute magnitude, $M_{UV}$. In accordance with current observational criterion, at each redshift $z \simeq 6-8$, resolved simulated galaxies with $M_{UV}\leq -17$ are identified as LBGs.

% ************************************************************************************************
\section{Comparing the simulations with LBG observations}
\label{sec3}
% *************************************************************************************************
Once we have identified the simulated galaxies that would be detected as LBGs in the snapshots at $z \simeq 6-8$, we compare their UV LFs, stellar mass functions, stellar mass densities (SMD) and specific star formation rates (sSFR) to the observed values in order to validate the simulations used in this work.
In this section we show a comparison between the intrinsic and dust-attenuated (observed) theoretical UV LFs and the data. The theoretical stellar mass functions, SMD and sSFR, and their comparison with observed values are shown in Appendix \ref{a2}.

We calculate the intrinsic UV LFs by binning simulated LBGs on the basis of their intrinsic (i.e. dust-unattenuated) UV magnitudes and dividing this by the width of the UV bin (0.5 dex), and the volume of the box. We find that the intrinsic UV LF shifts towards brighter luminosities and higher number densities with decreasing redshift from $z \simeq 8$ to $6$, as expected from the hierarchical structure formation scenario where successively larger systems build up with time from the merger of smaller systems. As seen in Fig. \ref{fig_uvlf}, we find that the simulated intrinsic LBG UV LFs are over-estimated with respect to the data, with the over-estimation increasing with decreasing redshift, hinting at the increasing dust enrichment of these galaxies; we note that the intrinsic UV LF at $z \simeq 8$ is already in agreement with the observations, requiring no dust correction at this redshift.

We then use the observed (i.e. dust attenuated) UV luminosity obtained for each galaxy at $z \simeq 6-8$ (see Secs. \ref{dust} and \ref{identify_lbgs}) to build the observed UV LF. It is encouraging to note that both the slope and the amplitude of the dust attenuated simulated LFs are in agreement with the observations for $z \simeq 6$ and $7$ as seen from Fig. \ref{fig_uvlf}; as mentioned before, the simulated UV LF at $z \simeq 8$ requires no dust to match the observations. As can be seen from the same figure, the effects of dust on continuum photons at $z \simeq 6,7$ are most severe for the most massive/luminous galaxies; indeed while $f_c \simeq 0.8$ for galaxies in halo masses of $M_h \leq 10^{10}\Msun$, the value drops steadily thereafter such that $f_c \simeq 0.01$ for the largest galaxies at both $z \simeq 6,7$. Further, as a result of galaxies typically being less massive, younger, and hence less dust enriched with increasing redshift, averaged over all LBGs $f_c$ drops from $\simeq 1$ at $z \simeq 8$ to $0.6$ at $z \simeq 7$ and $0.5$ at $z \simeq 6$. We briefly digress to note that these average $f_c$ values are about a factor of two higher than the values inferred in \citet{dayal2010a}. This is due to the different feedback models implemented in these two simulations: while only 25\% of the SN energy was used to power outflows in the simulation used in \citet{dayal2010a}, 50\% of the SN energy has been used to power outflows in the simulation used in this work. As a result of the much larger energy inputs that power SN winds in driving out gas from the galaxy, the typical stellar masses and SFRs obtained from the simulation used in this work are about a factor two lower than those presented in \citet{dayal2010a}.

Returning to our discussion regarding the observed LBG UV LFs, we notice that these also shift to progressively higher luminosities and/or higher number densities with decreasing redshift. \citet{dayal2013} have shown that this evolution depends on the luminosity range probed: the steady brightening of the bright end of the LF is driven by genuine physical luminosity evolution due to a fairly steady increase in the UV luminosity (and hence star-formation rates) in the most massive LBGs; the evolution at the faint end arises due to a mixture of both positive and negative luminosity and density evolution as these putative systems brighten and fade, and continuously form and merge into larger systems.

Finally, we find that the best fit Schechter parameters to the dust attenuated simulated UV LFs are: faint end slope $\alpha = (-1.9\pm0.2, -2.0\pm0.2, -1.8\pm0.2)$ and the knee of the luminosity function $M_{UV,*} =(-19.8\pm0.3, -19.7\pm0.2, -19.9\pm0.3) $ at $z \simeq (6,7,8)$. These values are in agreement with the values $\alpha = (-1.71\pm0.11, -1.90\pm 0.14, -2.02 \pm 0.22)$, $M_{UV,*} = (-20.04\pm 0.12, -19.9 \pm 0.2, -20.12 \pm 0.37)$ and $\alpha = (-1.74\pm0.16, -2.01\pm 0.21, -1.91 \pm 0.32)$, $M_{UV,*} = (-20.24\pm 0.19, -20.14 \pm 0.26, -20.1 \pm 0.52)$ inferred observationally by \citet{mclure2009, mclure2013} and \citet{bouwens2011b} at $z \simeq (6,7,8)$, respectively.

% ************************************************************************************************
\section{Simulating LAEs}
\label{sec4}
% *************************************************************************************************
In the previous section (and Appendix A), we have validated that the simulated galaxy population at $z \simeq 6-8$ is in agreement with a number of high-$z$ LBG observations. We now use these galaxy populations to identify the simulated galaxies that could be detected as LAEs. We start by describing the radiative transfer code (pCRASH) used to obtain reionization topologies for $\fesc=0.05,0.25,0.5,0.75,0.95$, which are utilized to calculate the IGM Ly$\alpha$ transmission. We then describe how these transmission values, and the effects of dust on Ly$\alpha$ photons are taken into account, so as to obtain the observed Ly$\alpha$ luminosity from the intrinsic value  for each galaxy. We note that all the radiative transfer and Ly$\alpha$ calculations are carried out using a single snapshot of the hydrodynamical simulation at the edge of the reionization epoch, at $z \simeq 6.7$.

\begin{figure*}
  \center{\includegraphics[width=1.0\textwidth]{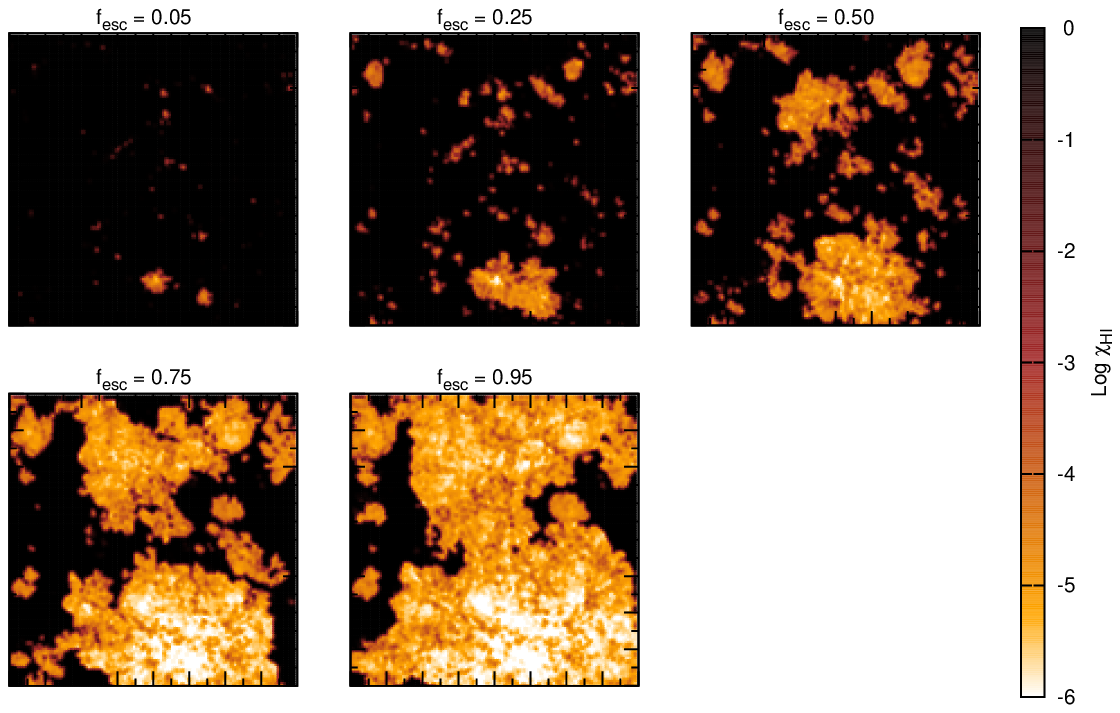}}
  \caption{Maps ($80 h^{-1}$ Mpc on a side) of the spatial distribution of \HI in a 2D cut through the center of the simulated box at $z \simeq 6.7$ obtained by running pCRASH for 50 Myrs using the $\fesc$ value marked above the panel. The colorbar shows the values (in log scale) of the \HI fraction. As can be seen, reionization proceeds faster with increasing $\fesc$ values. \label{fig_comp_fesc}}
\end{figure*}

% ************************************************************************************************
\subsection{Reionizing the Universe with pCRASH}
\label{pCRASH}
% *************************************************************************************************
As mentioned in Sec. \ref{intro}, the progress of reionization critically depends on the total number of \HI ionizing photons that can escape a galaxy and ionize the IGM around it. As expected, this depends both on the total number of \HI ionizing photons produced by a galaxy, as well as the fraction that can escape the galactic environment ($\fesc$). While the simulated star formation rates are in reasonable agreement with observations (see Appendix A), giving us a handle on the \HI ionizing photon production rate, the value of $\fesc$ remains only poorly understood: using a variety of theoretical models, the value of $\fesc$ has been found to range between $0.01-0.8$ \citep[see e.g.][]{ricotti-shull2000, fujita2003, razoumov2006, wise2009}. 

\begin{figure*}
  \center{\includegraphics[width=1.0\textwidth]{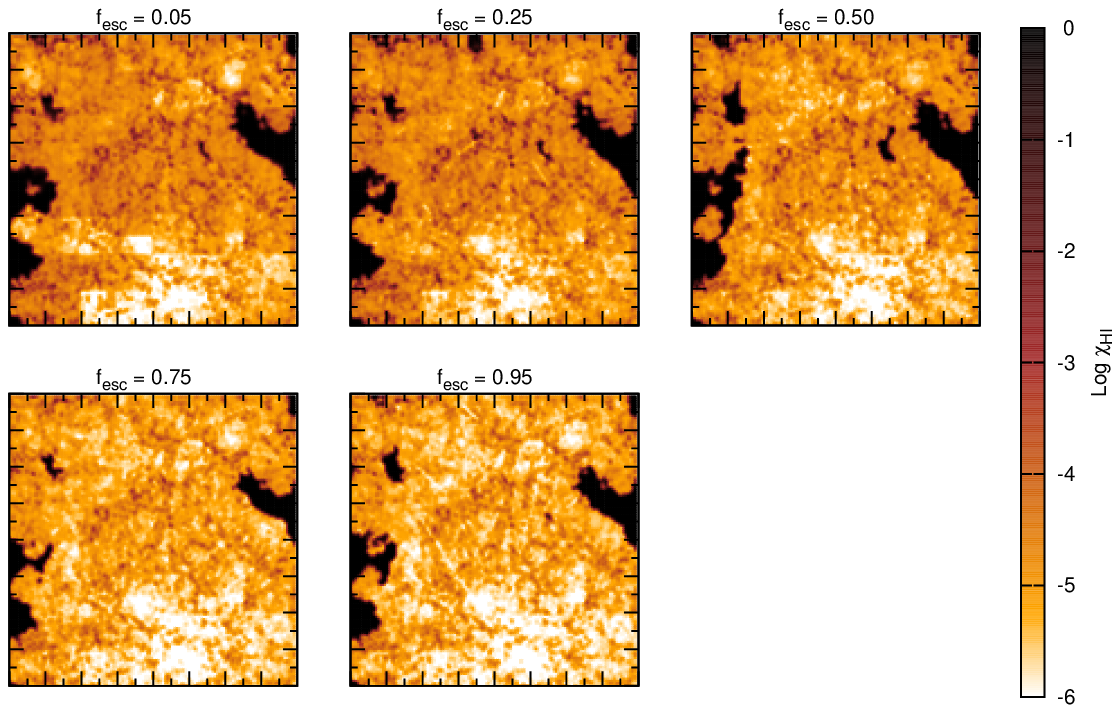}}
  \caption{Maps ($80 h^{-1}$ Mpc on a side) of the spatial distribution of \HI in a 2D cut through the simulated box at $z \simeq 6.7$ obtained by running pCRASH using the $\fesc$ value marked above the panel, until the average neutral hydrogen fraction drops to $\langle \chi_{HI} \rangle \simeq 0.1$. The colorbar shows the values (in log scale) of the \HI fraction. As seen, while the topology of reionization looks very similar for the same average \HI values, the degree of ionization close to any source increases with increasing $\fesc$.\label{fig_comp_fesc_XHI0.10}}
\end{figure*}

Due to the poor theoretical constraints available on the value of $\fesc$, we do not make any prior assumptions on its value. We instead post-process the hydrodynamical simulation snapshot at $z \simeq 6.7$ with the radiative transfer code pCRASH using five different values of $\fesc = 0.05$, $0.25$, $0.50$, $0.75$ and $0.95$ to study how varying $\fesc$ affects the progress of reionization. 

We now briefly describe pCRASH and interested readers are referred to \citet{ciardi2001}, \citet{maselli2003,maselli2009} and \citet{partl2011} for more details. pCRASH is a 3D grid-based MPI-parallelized radiative transfer code that utilizes a combination of ray tracing and Monte Carlo schemes. It follows the time evolution of the IGM gas properties including ionization fractions (\HI and \HeI, \HeII) and the temperature. A large number of point sources with different spectra emit photon packages into the medium. The relevant matter-radiation interactions (photoionization, recombination, collisional ionization) as well as the heating/cooling of the IGM (collisional ionization cooling, recombination cooling, collisional excitation cooling, bremsstrahlung, Compton cooling/heating, adiabatic cooling, photoheating) are calculated. In addition, the code includes diffuse radiation following recombinations of ionized atoms. 
For the radiative transfer runs, we use the density and temperature fields obtained from the $z \simeq 6.7$ snapshot of the hydrodynamical simulation as the inputs for pCRASH. For each resolved source in the SPH simulation, the luminosity and spectrum were calculated by summing up the spectra of all their star particles, as described in section \ref{identify_lbgs}; we processed $31855$ resolved sources from the snapshot at $z \simeq 6.7$. The spectrum of each source was binned into 125 frequency intervals ranging from $3.21\times 10^{15}$Hz to $3.30\times 10^{16}$Hz. Each source emitted $10^7$ rays and we used $10^6$ timesteps per $500$ Myrs on a $128^3$ grid; we used the mean local clumping factor of each cell obtained from the underlying density distribution from the SPH simulation to account for the clumpiness of the IGM \citep{raicevic2011}. 
Depending on the escape fraction the runs were carried out on $64$ or $128$ processors on AIP clusters. Each simulation was run until the IGM global hydrogen fraction dropped to $\langle \chi_{HI} \rangle \lsim 10^{-4}$, so that the entire progress of reionization could be mapped.
 
We now discuss the ionization fields obtained by running pCRASH on the SPH simulation snapshot at $z \simeq 6.7$ for the five different $\fesc$ values mentioned above. Firstly, the volume of the ionized region ($V_I$) carved out by any source depends on its total ionizing photon output such that
\begin{eqnarray}
V_{I}&\propto&\frac{Q \fesc}{\chi_{HI} n_H},
\label{eq_vel_ionfront}
\end{eqnarray}
where $Q$ is the total number of \HI ionizing photons produced by the source, $\chi_{HI}$ is the \HI fraction and $n_H$ represents the number density of hydrogen atoms. From this expression, it is clear that for a given source surrounded by an IGM of a given \HI density,  $V_I$ increases with $\fesc$. In other words, given a galaxy population, at any given time, the IGM is more ionized (i.e. reionization proceeds faster) for larger $\fesc$ values. 

Secondly, the photoionization rate ($\Gamma$) at a distance $r$ from a source depends on the source \HI photon emissivity ($L_\lambda \fesc$) such that
\begin{equation}
 \Gamma(r)= \int_0^{\lambda_L} \frac{L_\lambda\, \fesc}{4 \pi r^2} \sigma_L \bigg(\frac{\lambda}{\lambda_L}\bigg)^3 \frac{\lambda}{h c} d\lambda,
\label{eq_photoion_rate}
\end{equation}
where $L_\lambda$ is the total specific ionizing luminosity of the emitter, $\lambda_L$ is the Lyman limit wavelength (912 \AA), $\sigma_L$ is the hydrogen photoionization cross-section at $\lambda=\lambda_L$, $h$ is the Planck constant and $c$ represents the speed of light. Assuming ionization-recombination balance, it is this photoionization rate that determines the level of ionization, $\chi_{HI}$, around an emitter. Hence, for a given emitter, $\chi_{HI}$ is expected to decrease for increasing values of $\fesc$ \citep[see also Sec. 2.4,][]{dayal2008}.

These two effects can be clearly seen in Figs. \ref{fig_comp_fesc} and \ref{fig_comp_fesc_XHI0.10}: the former and latter show the ionization fields obtained for different $\fesc$ by running pCRASH for a given time (50 Myrs) and until the IGM is ionized to a level of $\langle \chi_{HI} \rangle \simeq 0.1$, respectively. From Fig. \ref{fig_comp_fesc}, it can be clearly seen that reionization proceeds faster for increasing $\fesc$ values: after 50 Myrs of running pCRASH, while the largest ionized region built has a size of about $15$ cMpc for $\fesc=0.05$, the entire box is almost reionized for $\fesc=0.95$. Indeed, the average value of $\chi_{HI}$ drops steadily with increasing $\fesc$ such that $\langle \chi_{HI} \rangle$ $= (0.97, 0.82, 0.62, 0.37, 0.20)$ for $\fesc=(0.05,0.25,0.5,0.75,0.95)$ respectively. Increasing $\fesc$ by a factor of $19$, from 0.05 to 0.95 has the effect of decreasing the neutral hydrogen fraction; while the IGM is essentially neutral for $\fesc=0.05$, it is almost ionized for $\fesc=0.95$.

From Fig. \ref{fig_comp_fesc_XHI0.10}, we find that the topology of reionization looks similar for all the five $\fesc$ values by the time the average \HI fraction has dropped to $\langle \chi_{HI} \rangle =0.1$. However, the time taken to reach this average ionization state is very different for the five runs; while the run with $\fesc=0.95$ took $80\ {\rm Myrs}$ to reach this ionization level, the runs with $\fesc=0.05$ took about ten times longer ($\simeq 800\ {\rm Myrs}$) to build up these ionized regions. We also note that although the spatial distribution of the ionization fields looks very similar for the varying $\fesc$ values, the level of ionization in any given cell increases with increasing $\fesc$ as seen from the same figure (see also Eqn. \ref{eq_photoion_rate}).

To summarize, we find that increasing $\fesc$ both accelerates the progress of reionization and leads to higher ionization fractions in the ionized regions. A combination of both these factors leads to a larger IGM  Ly$\alpha$ transmission $T_{\alpha}$ with increasing $\fesc$, as explained in the following.

% ************************************************************************************************
\subsection{Identifying LAEs}
\label{identify_laes}
% *************************************************************************************************
In this section, we start by explaining how we calculate the intrinsic Ly$\alpha$ luminosity produced by each simulated galaxy at $z \simeq 6.7$. We then show how we calculate the \Lya attenuation by ISM dust and IGM \HI to determine the fraction of this intrinsic luminosity that is finally observed.

% ************************************************************************************************
\subsubsection{Intrinsic Ly$\alpha$ luminosity}
\label{intr_lyalpha}
% *************************************************************************************************
As mentioned above, star formation in galaxies produces photons more energetic than 1 Ryd, with a certain fraction ($\fesc$) escaping into, and ionizing the IGM. The rest of these photons $(1-\fesc)$ ionize the \HI in the ISM. Due to the high density of the ISM, these electrons and protons recombine on very short timescales giving rise to a Ly$\alpha$ emission line of luminosity
\begin{eqnarray}
L_{\alpha}^{int} &=& \frac{2}{3}Q(1-\fesc)h\nu_{\alpha},
\label{eq_lum_alpha_intr}
\end{eqnarray}
where the factor two-thirds arises due to our assumption of case-B recombinations for optically thick \HI in the ISM \citep{osterbrock1989}, and $\nu_{\alpha}$ represents the \Lya frequency in the galaxy rest frame.

However, this line is Doppler broadened by the rotation of the galaxy so that the complete line profile can be expressed as
\begin{equation}
L_{\alpha}^{int}(\nu) = \frac{2}{3}Q(1-\fesc) h\nu_{\alpha}\frac{1}{\sqrt{\pi}\Delta\nu_d} \exp \left[-\frac{(\nu-\nu_{\alpha})^2}{\Delta\nu_d^2}\right],
\end{equation}
where  $\Delta \nu_d = (v_c/c)\nu_\alpha$. For realistic halo and disk properties, the rotation velocity of the galaxy ($v_c$) can range between $v_h$ and $2v_h$, where $v_h$ is the halo rotation velocity \citep{Mo1998,Cole2000}. We use the central value of $v_c=1.5 v_h$ in all the calculations presented in this paper. 

% ************************************************************************************************
\subsubsection{Effects of dust}
\label{dust_lyalpha}
% *************************************************************************************************
The intrinsic Ly$\alpha$ luminosity produced by a galaxy has to penetrate through dust in the ISM, with only a fraction $f_\alpha$ emerging out of the galactic environment. In Sec. \ref{dust}, we have shown calculations for the total dust enrichment of each simulated galaxy and the resulting value of $f_c$. The relation between $f_\alpha$ and $f_c$ depends on the adopted extinction curve if dust is homogeneously distributed; it also depends on the differential effects of radiative transfer on \Lya and UV photons if dust is inhomogeneously distributed/clumped \citep{neufeld1991, hansen-oh2006}. While some pieces of evidence exist that the SN extinction curve \citep{bianchi-schneider2007} can successfully be used to interpret the observed properties of the most distant quasars \citep{maiolino2006} and gamma-ray bursts \citep{stratta2007}, the effect of dust inhomogeneities in enhancing the escape fraction of Ly$\alpha$ photons with respect to continuum photons through the Neufeld effect \citep{neufeld1991} remains controversial. On the one hand, \citet{finkelstein2009} have found tentative observational evidence of $f_\alpha/f_c>1$ for $z \simeq 4.5$ LAEs and \citet{dayal2009,dayal2011} have shown that models require $f_\alpha/f_c>1$ to reproduce LAE data at $z \lsim 6$. On the other hand, \citet{laursen2013} have shown that a value of $f_\alpha>f_c$ requires very special circumstances (no bulk outflows, very high metallicity, very high density of the warm neutral medium and a low density and highly ionized medium) that are unlikely to exist in any realistic ISM; moreover, they show that a value of $f_\alpha/f_c>1$ results in very narrow Ly$\alpha$ lines that are sensitive to infalling gas, resulting in low $T_\alpha$ values. Due to its poor understanding, the relative escape fraction $f_\alpha/f_c$ is left as a free parameter in our model and its value is fixed by matching the theoretical LAE \Lya and UV LFs to the observations, as shown in Sec. \ref{sec5} that follows.

% ************************************************************************************************
\subsubsection{Effects of the IGM}
\label{IGM_lyalpha}
% *************************************************************************************************
The \Lya photons that escape out of a galaxy unabsorbed by dust are attenuated by the \HI they encounter along the line of sight (LOS) between the emitter and the observer, with a fraction $T_\alpha = e^{-\tau_{\alpha}}$ being transmitted through the IGM. This optical depth to \HI ($\tau_\alpha$) can be calculated as
\begin{eqnarray}
\tau_{\alpha} (v)&=&\int_{r_{em}}^{r_{obs}} \sigma_0\ \Phi\left(v+v_P\left(r\right)\right)\ n_{HI}(r)\ \mathrm{d}r,
\label{tau_alpha}
\end{eqnarray}
where $v=(\lambda-\lambda_{\alpha})\lambda_{\alpha}^{-1}c$ is the rest-frame velocity of a photon with frequency $\lambda$ relative to the Ly$\alpha$ line center at $\lambda_{\alpha}=1216$\AA. Further, $v_P$ accounts for the IGM peculiar velocity, $n_{HI}(r)$ the \HI density at a physical distance $r$ from the emitter, $\sigma_0 = \pi e^2 f/m_e c$ is the absorption cross section, $f$ is the oscillator strength, $e$ is the electron charge, $m_e$ is the electron mass,  and $\Phi(v)$ is the Voigt profile; this profile consists of a Gaussian core and Lorentzian damping wings. For regions of low \HI density, pressure line broadening becomes unimportant and the line profile can be approximated by the Gaussian core. However, the Lorentzian damping wings become important in regions of high \HI density and the complete profile must then be used. Although it is computationally quite expensive, we use the complete Voigt profile for all our calculations. We note that we include the effects of peculiar velocities ($v_P$) in determining $\tau_\alpha$: inflows (outflows) of gas towards (from) a galaxy will blueshift (redshift) the Ly$\alpha$ photons, increasing (decreasing) $\tau_{\alpha}$, which leads to a corresponding decrease (increase) in $T_{\alpha}$. 
We compute $\tau_{\alpha}$ by stepping through our simulation box which is divided into a grid of $128^3$ cells, with each cell having a size of $117$ physical kpc; this is the same grid that is used to run pCRASH, as described in Sec. \ref{pCRASH}. For any galaxy, we start calculating $\tau_{\alpha}$ at the galaxy position $r_{em}=0$, until the edge of the box is reached $r_{obs}$. The value of $v_P(r)$ and $n_{HI}(r)$ in each cell are then obtained from GADGET-2 and pCRASH runs, respectively. A single GADGET-2 snapshot at $z\simeq6.7$ suffices for our work because the \Lya line redshifts out of resonance with \HI extremely quickly; to a first approximation, the spatial scale imposed by the Gunn-Peterson damping wing on the size of the \HII region corresponds to a separation of about $280$ kpc (physical) at $z = 6$ \citep{miralda-escude1998}. Interested readers are referred to \citet{dayal2011} for complete details of this calculation.

We use the ionization fields obtained by running pCRASH (see Sec. \ref{pCRASH}) with the five different $\fesc$ values to calculate the average $T_\alpha$ for each galaxy along 48 lines of sight (LOS). For a given ionization state, the final $T_\alpha$ value of each galaxy is taken to be the average along all 48 LOS.
For each galaxy we compute the observed Ly$\alpha$ luminosity ($L_{\alpha}^{obs}$) by using the individual values for $f_{\alpha}$ and $T_{\alpha}$ of the respective galaxy.
\begin{equation}
L_{\alpha}^{obs} = f_{\alpha} T_{\alpha} L_{\alpha}^{int}.
\label{eq_lum_alpha_obs}
\end{equation}
In consistency with current observational criteria, galaxies with $L_{\alpha}^{obs}\ge10^{42}$erg s$^{-1}$ and an equivalent width ($EW = L_\alpha^{obs}/L_c^{obs} \geq 20$\AA) are then identified as LAEs and used to build the LAE UV and Ly$\alpha$ LFs. We note that the \Lya LFs along different LOS are subject to cosmic variance as shown in Appendix \ref{a1}.

We also remind the reader that $T_\alpha$ is fixed from the ionization state of the IGM as explained above. Then, the only free parameter that we are left with to identify LAEs is the relative escape fraction of Ly$\alpha$ and continuum photons i.e. $f_\alpha/f_c$. This final free parameter is fixed by matching the simulated LFs to the observed ones, as explained in Sec. \ref{sec5}.

\begin{figure}
  \center{\includegraphics[width=0.5\textwidth]{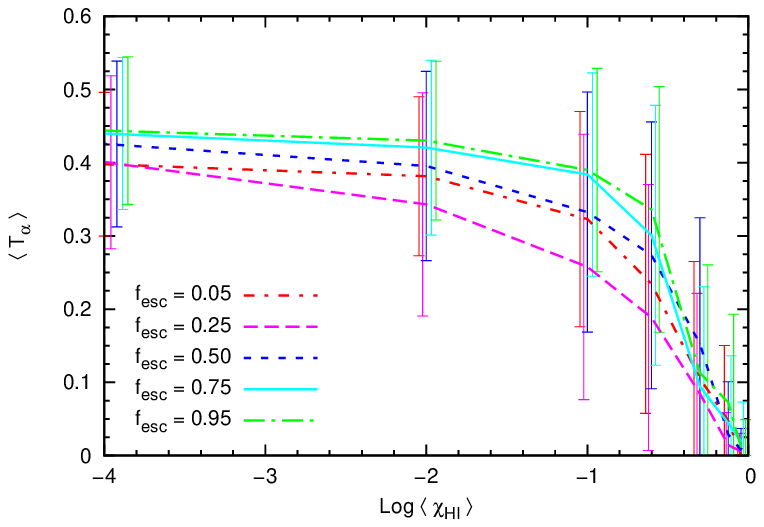}}
  \caption{Mean transmission of all galaxies (averaged over 48 LOS) in the simulation snapshot at $z \simeq 6.7$ as a function of the mean IGM hydrogen neutral fraction. Red (dash-dotted), magenta (long dashed), dark blue (short dashed), light blue (dotted) and green (long dashed-dotted) lines show the relation for $\fesc=0.05$, $0.25$, $0.5$, $0.75$ and $0.95$ and error bars indicate the 1$\sigma$ dispersion.\label{fig_sources_XHI_transmission}}
\end{figure}

Before proceeding to a comparison between the theoretical results and observational data, we briefly digress to show the relation between $T_\alpha$ and \avchi for our 5 $\fesc$ values. As can be seen from Fig. \ref{fig_sources_XHI_transmission} (see also Sec. \ref{pCRASH}), the mean IGM transmission ($\langle T_{\alpha} \rangle$) averaged over all galaxies increases with decreasing \avchi values: as ionized regions grow larger, they allow all galaxies to transmit more of their Ly$\alpha$ flux through the IGM. The inhomogeneity of the ionized regions also leads to a huge variation in $T_\alpha$ as seen from the $1\sigma$ dispersions in the same figure. Finally, our mean IGM transmission values are consistent with the results obtained by \citet{jensen2013}, \citet{iliev2008} and \citet{dijkstra2007} for $\fesc=0.05$; we note that Ly$\alpha$ transmission calculations have not been performed for escape fractions higher than this value.

% ***************************************************************************
\section{Joint constraints on reionization topology and dust} 
\label{sec5}
% ****************************************************************************
From Secs. \ref{pCRASH} and \ref{identify_laes}, it is clear that $\fesc$ affects the observed Ly$\alpha$ luminosity both through its effect on the intrinsic Ly$\alpha$ luminosity and through its effect on the ionization state of the IGM. Further, for a given $\fesc$ and ionization state, the only free parameter left in our model is the escape fraction of \Lya photons compared to UV-continuum photons, $f_\alpha/f_c$. In this section we use the LAE data accumulated by \citet{kashikawa2011} to simultaneously constrain and find best fits to our 3 parameters, $\fesc$, $\langle \chi_{HI} \rangle $ and $f_\alpha/f_c$. We use pCRASH outputs for $\langle \chi_{HI} \rangle  \simeq 0.9,0.75,0.5,0.25,0.1,0.01$ and $10^{-4}$ for each of the five $\fesc=0.05,0.25,0.5,0.75,0.95$ at $z \simeq 6.7$. 
For each such combination, we start from the special case of assuming $f_\alpha/f_c=0.68$ which corresponds to dust being homogeneously distributed in the ISM in Sec. \ref{homogenous_dust}, and progress to the more general scenario of varying $f_\alpha/f_c$ to best fit the observations, in Sec. \ref{clumped_dust}.

% ***************************************************************************
\subsection{A special case: homogeneous dust} 
\label{homogenous_dust}
% ***************************************************************************

\begin{figure*}
  \center{\includegraphics[width=1.0\textwidth]{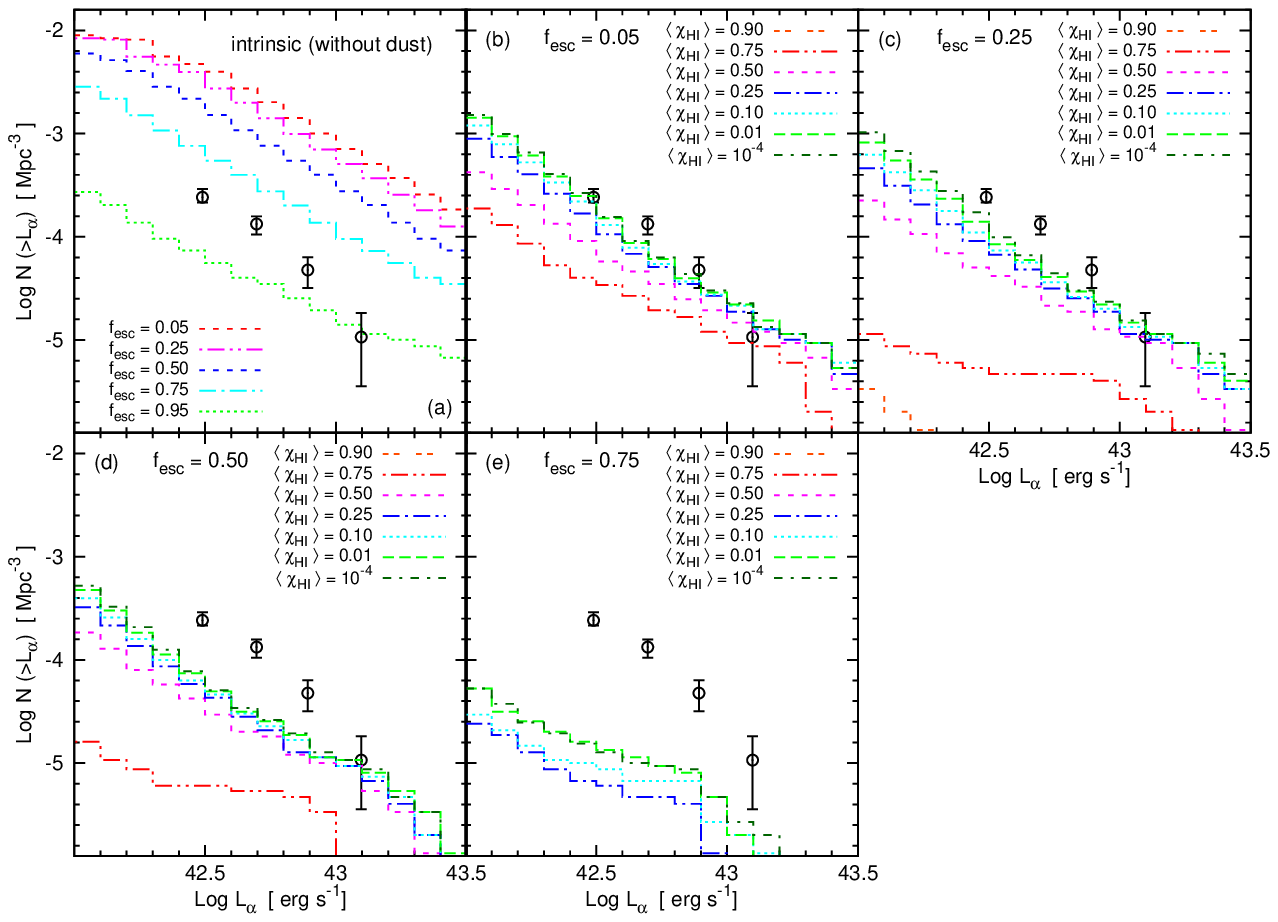}}
  \caption{The cumulative \Lya LFs. Panel (a) shows the intrinsic \Lya LFs for escape fractions $\fesc=0.05$, $0.25$, $0.50$, $0.75$, as marked. The lines in the other panels show the results for homogeneously distributed SNII dust ($f_\alpha/f_c=0.68$) for different \avchi states ($\langle \chi_{HI} \rangle \simeq 0.90$, $0.75$, $0.50$, $0.25$, $0.10$, $0.01$, $10^{-4}$) obtained using the fixed $\fesc$ value marked above the panel. In each panel, black open circles show the \Lya LFs at $z\simeq6.5$ inferred observationally by \citet{kashikawa2011}. Due to the EW selection criterion no LAEs are identified for $\fesc=0.95$.}
\label{fig_LAE}
\end{figure*}

As shown in Eqn. \ref{eq_lum_alpha_intr}, the intrinsic Ly$\alpha$ luminosity produced by a galaxy depends on the fraction of \HI ionizing photons that ionize the \HI in the ISM. It is thus naturally expected that $L_\alpha^{int}$ decreases with increasing $\fesc$ since this leads to fewer photons being available for ionization in the ISM. This behaviour can be seen from panel (a) of Fig. \ref{fig_LAE} where we find that the amplitude of the \Lya LF drops by a factor of about 40 as $\fesc$ increases from $0.05$ to $0.95$. This drop seems surprising at the first glance: as $\fesc$ increases by a factor of about 19, the intrinsic \Lya LF would be expected to drop by the same amount. However, this result can be easily explained by the fact that we identify galaxies as LAEs based on the \Lya luminosity being larger than $10^{42}$erg s$^{-1}$ and an $EW \geq 20$\AA. As $\fesc$ increases, fewer and fewer galaxies are able to fulfil these selection criteria, leading to an enhanced drop in the amplitude of the \Lya LF.

\begin{figure*}
  \center{\includegraphics[width=0.8\textwidth]{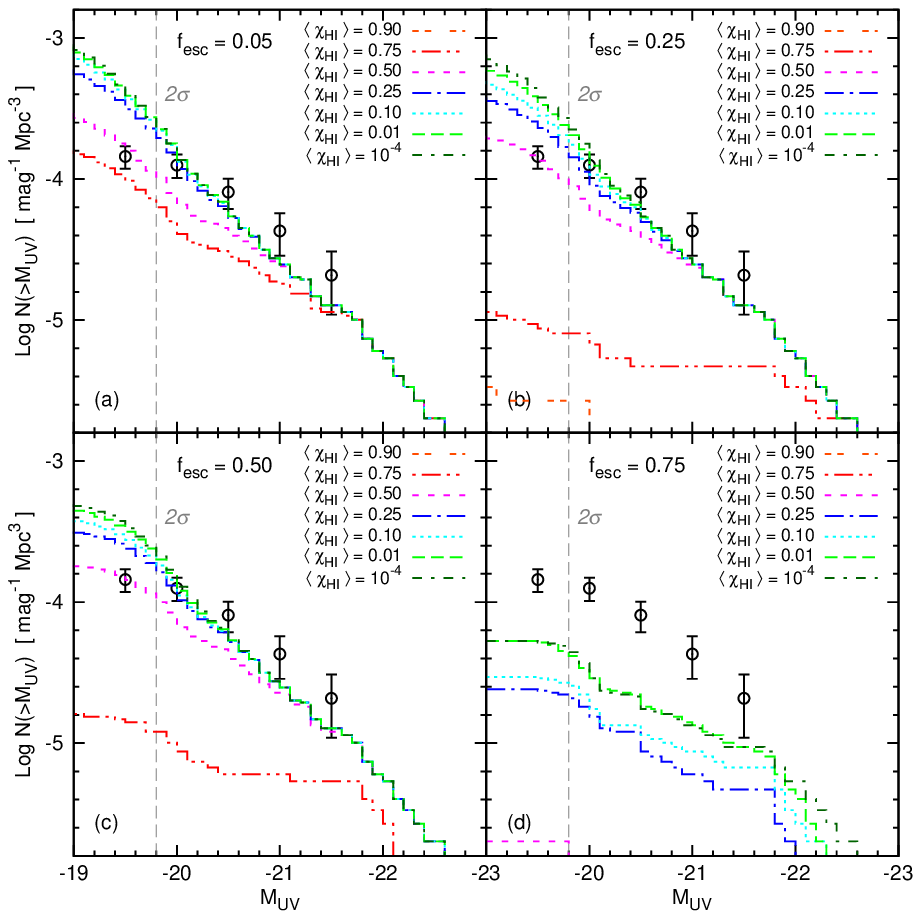}}
  \caption{The cumulative UV LFs for simulated LAEs using a homogeneous dust case ($f_{\alpha}/f_c=0.68$). The lines in the panels show the UV LFs for different \avchi states ($\langle \chi_{HI} \rangle \simeq 0.90$, $0.75$, $0.50$, $0.25$, $0.10$, $0.01$, $10^{-4}$) obtained using the fixed $\fesc$ value marked above the panel. In each panel, black open circles show the LAE UV LFs at $z\simeq6.5$ inferred observationally by \citet{kashikawa2011}. The vertical grey dotted lines indicate the limiting $2\sigma$ magnitude in the z' band \citep[see][]{kashikawa2011}; measurements for fainter magnitudes may be uncertain because the corresponding z'-band magnitudes are no longer reliable. Due to the EW selection criterion no LAEs are identified for $\fesc=0.95$.}
  \label{fig_LAE_UV}
\end{figure*}

The ratio $f_{\alpha}/f_c$ depends on many physical effects \citep[see e.g.][]{laursen2013}, including the distribution of the Ly$\alpha$ sources in the dusty ISM \citep{scarlata2009}, the ISM \HI column densities and velocities. However, a natural consequence of simulating cosmological volumes is that we are unable to resolve the ISM of individual galaxies. We therefore start with the special scenario wherein dust is homogeneously distributed in the ISM and Ly$\alpha$ photons are not scattered by dust and \HI within the ISM. Since SNII are assumed to be the main sources of dust in our model, we deduce the ratio $f_{\alpha}/f_c=0.68$ purely from the corresponding SN extinction curve \citep{bianchi-schneider2007}.
This fixed ratio is then applied uniformly to all galaxies in the simulation snapshot at $z \simeq 6.7$. Once $f_\alpha$ has been fixed using this ratio, the only two parameters that can affect the slope and amplitude of the \Lya LFs are $\fesc$ and $\chi_{HI}$.

We start by discussing the \Lya LF evolution as a function of $\langle \chi_{HI} \rangle$. As expected from Eqn. \ref{tau_alpha}, for a given $\fesc$, $\tau_\alpha$ decreases with decreasing $\langle \chi_{HI} \rangle$, leading to an increase in the IGM \Lya transmission. This naturally results in an increase in the amplitude of the \Lya LF by making galaxies more luminous in the observed \Lya luminosity, $L_\alpha^{obs}$, as can be seen from panels (b)-(e) of Fig. \ref{fig_LAE}. However, the effect of decreasing $\chi_{HI}$ is the strongest on the visibility of the faintest galaxies and decreases with increasing luminosity. This is because the spatial scale imposed by the Gunn-Peterson damping wing on the size of the \HII region corresponds to a separation of about 280 kpc (physical) at $z \simeq 7$ \citep{miralda-escude1998}. The most massive galaxies are easily able to carve out \HII regions of this size as a result of their relatively large \HI ionizing photon output. However, in the initial stages of reionization, only those faint galaxies that are clustered are able to build \HII regions of this size and become visible as LAEs \citep{mcquinn2007, dayal2009}. Hence, for a given $\fesc$, the faint-end slope of the \Lya LF becomes steeper with decreasing $\langle \chi_{HI} \rangle$ values as seen from panels (b)-(e) of Fig. \ref{fig_LAE}. We also see that the slope and amplitude of the \Lya LFs start converging for $\langle \chi_{HI} \rangle \simeq 10^{-2}$; this corresponds to the average \HI fraction at which galaxies typically inhabit \HII regions such that the red part of the \Lya line escapes unattenuated by \HI. 

We now discuss the evolution of the \Lya LFs with $\fesc$. As we have already noted in Fig. \ref{fig_comp_fesc_XHI0.10}, although the spatial distribution of the ionization fields looks very similar for the varying $\fesc$ values for a given \avchi, the $\chi_{HI}$ value in any given cell decreases with increasing $\fesc$. This leads to an increase in $T_\alpha$ with $\fesc$ such that for \avchi $\simeq 0.1$, averaged over all LAEs, $\left<T_{\alpha}\right>=(0.43, 0.45, 0.48, 0.59)$ for $\fesc=(0.05, 0.25, 0.50, 0.75)$. However, this slight increase in $T_\alpha$ is compensated by the decrease in the intrinsic \Lya luminosity with increasing $\fesc$ as shown in panel (a) of Fig. \ref{fig_LAE}. As a result, the amplitude of the resulting \Lya LFs decreases slightly as $\fesc$ increases from 0.05 to 0.5 for a given \avchi; however, the slopes are very similar since $\fesc$ affects the intrinsic luminosity of all simulated galaxies by the same factor. The \Lya LF drops steeply between $\fesc=0.5$ and 0.75 and there are no simulated galaxies that would be identified as LAEs for $\fesc=0.95$. This steep drop for $\fesc>0.5$ arises due to our imposed criterion of the observed \Lya EW being larger than 20\AA. For a continually star-forming galaxy with an age of about 200Myr, a stellar metallicity of $Z_* = 0.02Z_\odot$ and $\fesc=0$, the intrinsic \Lya EW has a value of about $114$\AA; we note that this intrinsic EW would decrease for increasing values of $Z_*$ and $\fesc$. For a value of $\fesc=(0.75,\ 0.95)$, this intrinsic EW then reduces to about (28.5, 5.7)\AA\,. Then, taking into account the effect of the IGM transmission very few (none) of our simulated galaxies fulfil the EW selection criterion for $\fesc=0.75$ (0.95) leading to a steep drop in the number density of LAEs; the total number density of LAEs drops to zero for $\fesc=0.95$.

Finally, we note that both the slope and amplitude of the simulated \Lya LFs are in agreement with observations for $\fesc =0.05$ and \avchi$\leq 0.1$; for $\fesc\geq 0.25$, the simulated \Lya LFs are underestimated with respect to the data, even for a fully ionized IGM.

\begin{figure*}
  \center{\includegraphics[width=1.0\textwidth]{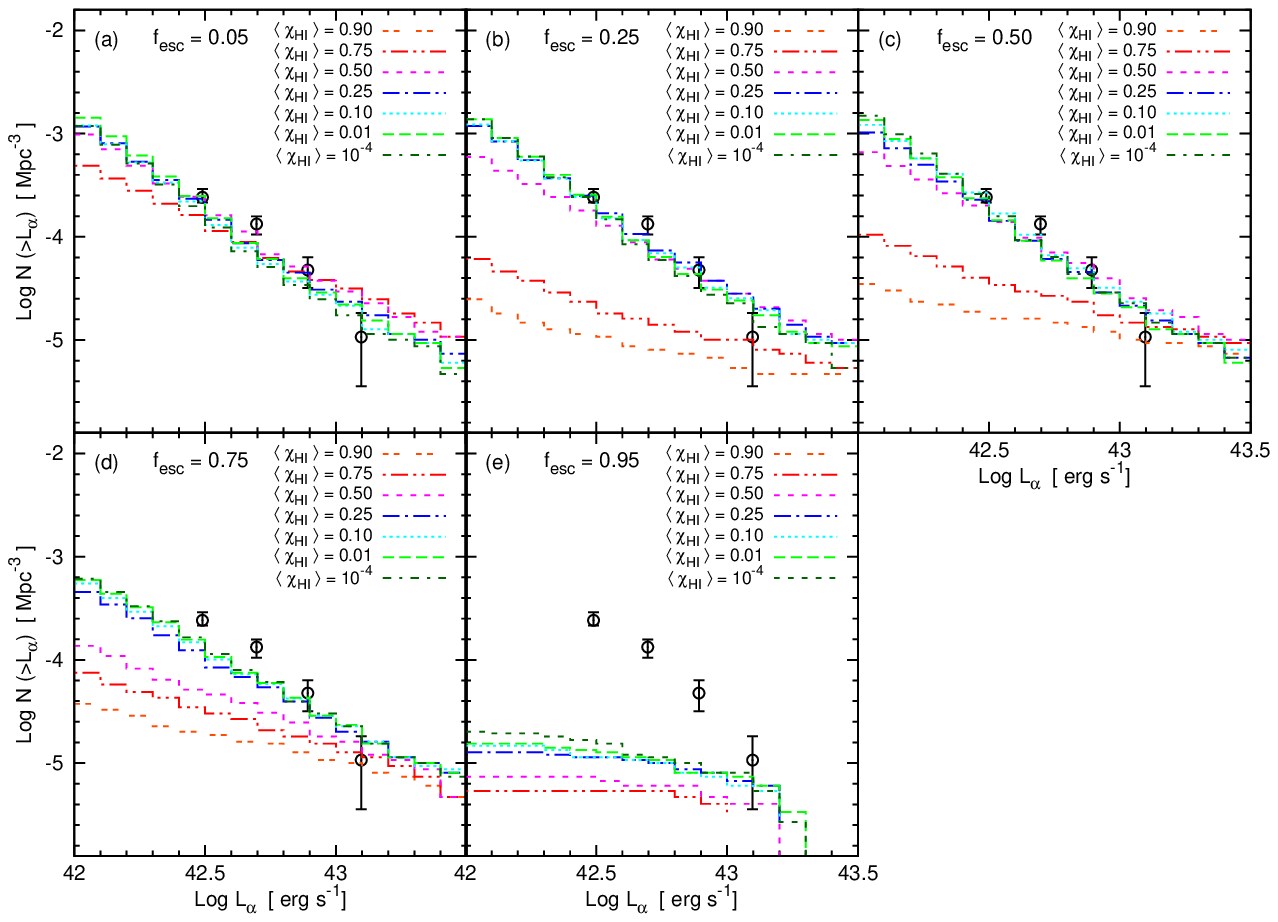}}
  \caption{The cumulative \Lya LFs. The panels show the best-fit simulated results for the $f_\alpha/f_c$ values shown in Table \ref{table1}. The panels show the results for different \avchi states ($\langle \chi_{HI} \rangle \simeq 0.90$, $0.75$, $0.50$, $0.25$, $0.10$, $0.01$, $10^{-4}$) obtained using the fixed $\fesc$ value marked above the panel. In each panel, black open circles show the \Lya LFs at $z\simeq6.5$ inferred observationally by \citet{kashikawa2011}.}
\label{fig_LAE_bestfit}
\end{figure*}

The UV LFs are not directly dependent on the IGM ionization state. However, both their slope and amplitude are affected by \avchi since these are the UV LFs for the simulated galaxies that would be identified as LAEs on the basis of $L_\alpha^{obs}$ and the EW. It is interesting to note that with our deeper \Lya luminosity limit of $L_\alpha^{obs}\geq 10^{42}$\lobs compared to the observational limit of $10^{42.4}$ \lobs, we find that the cumulative UV LFs keep rising until $M_{UV} \leq -19$ for $\fesc\leq 0.5$ as seen from panels (a)-(c) of Fig. \ref{fig_LAE_UV}. From these same panels, we find that the bright-end slope of the UV LF ($M_{UV} \leq -21.5$) matches the observed UV LFs for all values of \avchi and $\fesc\leq 0.5$. Further, the UV LFs start converging at much higher values of \avchi$\simeq 0.25$, compared to the value of $0.1$ required for the \Lya LFs to settle. Similar to the behaviour seen for the \Lya LF, for a given $\fesc\leq 0.5$, the faint end of the UV LF steepens with decreasing \avchi as more and more of these galaxies are able to transmit enough of their \Lya flux to be visible as LAEs. Again, for a given \avchi, while the amplitude of the UV LFs decreases slightly between $\fesc=0.05$ and 0.5, the slope remains the same. The steep drop in the number of LAEs for $\fesc\simeq 0.75$ naturally leads to a drop in the UV LF shown in panel (d) of the same figure; as for the \Lya LF, the UV LF amplitude is zero for $\fesc=0.95$ due to no simulated galaxies being classified as LAEs.

Finally, we note that by jointly reproducing the observed \Lya and UV LFs, we can constrain for $\fesc=0.05$ one of the most important parameters for reionization: \avchi$\leq 0.1$ or $\langle T_{\alpha} \rangle_{LAE} \geq 0.43$. 

However, we caution the reader that this result is only valid in case the ISM dust is homogeneously distributed. Since using only the UV LFs allows a much larger parameter range of \avchi$\leq 0.25$, a question that arises is whether this larger parameter space could also be reconciled with the \Lya LFs given clumped dust that could increase $L_\alpha^{obs}$ by increasing $f_\alpha$. We carry out these calculations in Sec. \ref{clumped_dust} that follows.

\begin{figure*}
  \center{\includegraphics[width=0.8\textwidth]{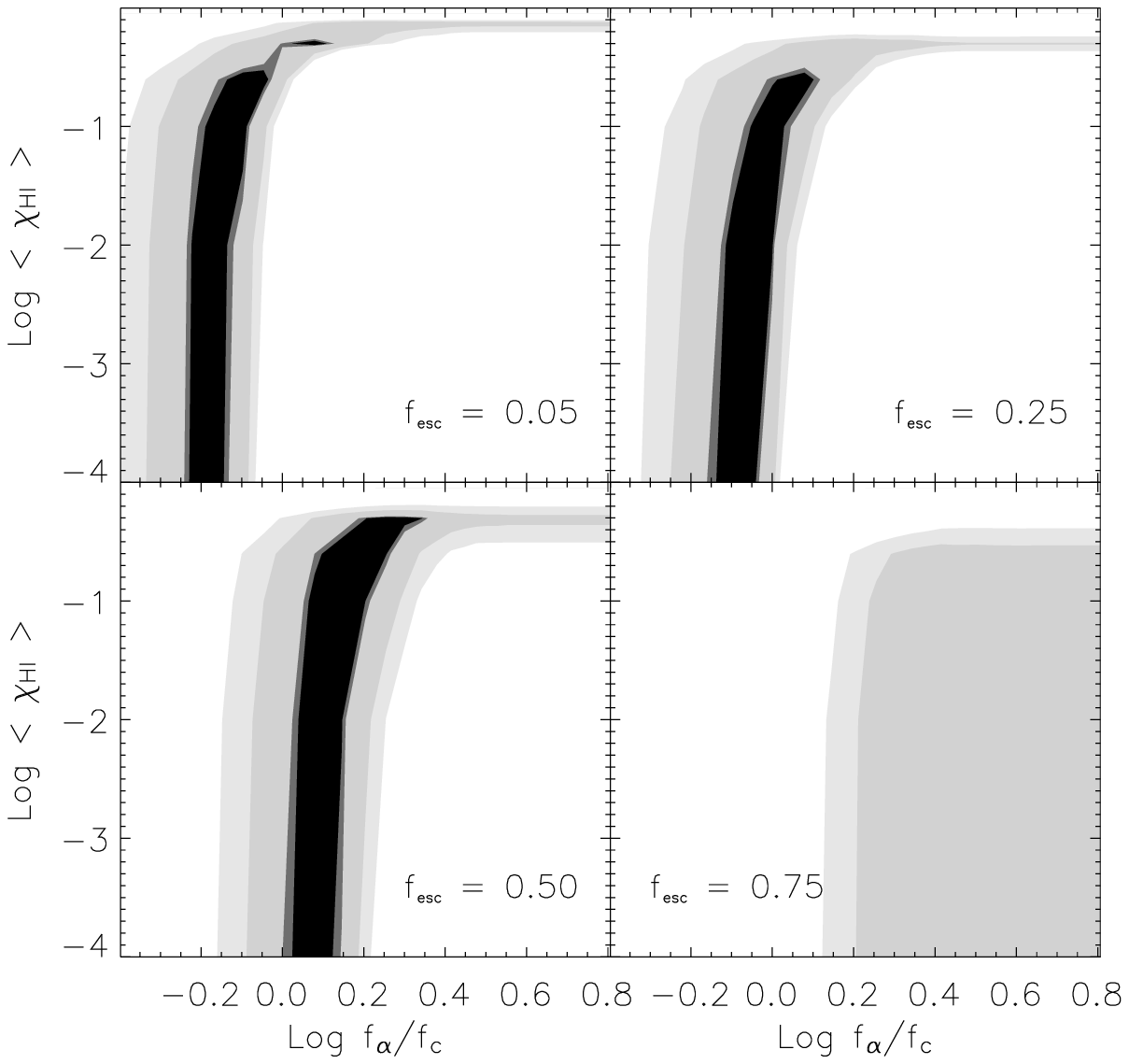}}
  \caption{$1-5\sigma$ probability contours (black to light gray respectively) for the combinations of \avchi, $\fesc$ and $f_\alpha/f_c$ shown in Table. \ref{table1} that best fit the observed \Lya LF data at $z\simeq6.5$ \citet{kashikawa2011}. Within a $1\sigma$ error, we can not distinguish between \avchi$\simeq 10^{-4}$ to $0.5$, $f_\alpha/f_c$ from $0.6$ to $1.8$ and $\fesc$ ranging between $0.05$ and $0.5$. There exists a {\it degeneracy} between the IGM ionization state, escape fraction of \HI ionizing photons and the dust clumping inside high-redshift galaxies such that a decrease in the IGM \Lya transmission can be compensated by an increase in the intrinsic \Lya luminosity produced, and the fraction of this luminosity that can escape the galaxy. }
\label{fig_chi3D}
\end{figure*}

\begin{figure*}
  \center{\includegraphics[width=0.8\textwidth]{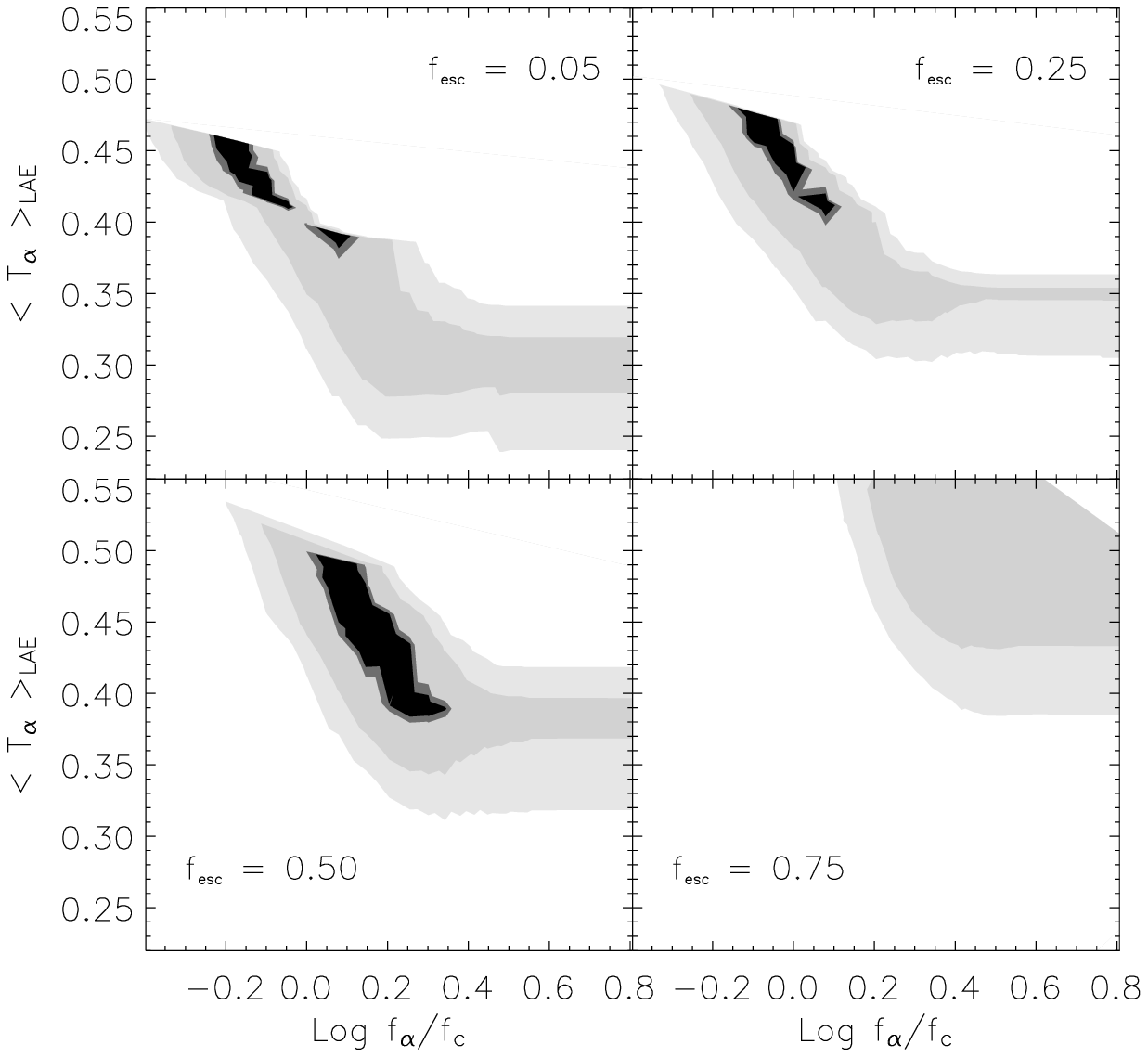}}
  \caption{$1-5\sigma$ probability contours (black to light gray respectively) for the combinations of \avchi, $\fesc$ and $f_\alpha/f_c$ shown in Table. \ref{table1} that best fit the observed \Lya LF data at $z\simeq6.5$ \citet{kashikawa2011}. Within a $1\sigma$ error, we can not distinguish between $\langle T_{\alpha}\rangle_{LAE}\simeq 0.38$ to $0.5$, $f_\alpha/f_c$ from $0.6$ to $1.8$ and $\fesc$ ranging between $0.05$ and $0.5$. There exists a {\it degeneracy} between the IGM Ly$\alpha$ transmission, escape fraction of \HI ionizing photons and the dust clumping inside high-redshift galaxies such that a decrease in the IGM \Lya transmission can be compensated by an increase in the intrinsic \Lya luminosity produced, and the fraction of this luminosity that can escape the galaxy. }
\label{fig_chi3D_transmission}
\end{figure*}

% ***************************************************************************
\subsection{Clumped dust} 
\label{clumped_dust}
% ****************************************************************************.
A number of works have shown that inhomogeneously distributed (clumped) dust in the ISM can enhance the escape fraction of Ly$\alpha$ photons compared to UV photons which are not resonantly scattered by \HI \citep{neufeld1991, hansen-oh2006}. However, the amount by which the ratio $f_\alpha/f_c$ can be enhanced by clumped dust sensitively depends on the amount of dust that is bound in cold neutral gas clouds, and the number of these encountered by \Lya photons within the galaxy. Since such sub-grid parameters can not be resolved by our cosmological simulation, we use $f_\alpha/f_c$ as a free parameter and explore the range required to reproduce the data for the different $\fesc$ and \avchi combinations discussed in Sec. \ref{homogenous_dust}. 

As shown in Section \ref{homogenous_dust}, an increasing $\fesc$ reduces the intrinsic Ly$\alpha$ luminosity for all galaxies, leading to a drop in the amplitude of the \Lya LF. However, since $L_{\alpha}^{obs}\propto(1-\fesc)f_{\alpha} T_\alpha$, a decrease in the intrinsic \Lya luminosity (due to an increasing $\fesc$) can be compensated by a larger fraction of \Lya photons escaping out of the galaxy (i.e. a higher $f_\alpha$) and/or a larger amount being transmitted through the IGM (i.e. a larger $T_\alpha$). Hence, for a given \avchi (i.e. $T_\alpha$) by allowing values of $f_\alpha/f_c>0.68$, the amplitude of the \Lya LF can be boosted. However, it is important to note that since we use a constant value of $f_\alpha/f_c$ for all galaxies, it only changes the amplitude of the \Lya LF at the bright end; raising this ratio makes the faint-end slope slightly steeper since a number of galaxies which are at our detection limit ($L_\alpha \geq 10^{42}$\lobs and $EW\geq 20$\AA) become visible as LAEs with increasing $f_\alpha/f_c$.

It is interesting to note that even using $f_\alpha/f_c$ ratios such that $f_\alpha$ saturates to 1, the simulated \Lya LFs are well below the observations for \avchi$\simeq 0.75$, for all  $\fesc = 0.05$ to $0.95$, as seen from panels (a)-(e) of Fig. \ref{fig_LAE_bestfit}. This is because the \HII regions built by faint galaxies are too small to allow a large IGM \Lya transmission, even for $\fesc=0.95$. We now discuss the results for varying $\fesc$: as mentioned in Sec. \ref{homogenous_dust}, the \Lya and UV LFs can be well reproduced using the homogeneous dust case for $\fesc=0.05$ for \avchi$\leq 0.1$. In the case of clumped dust, the simulated LFs match the observations for $\fesc=0.05$ for \avchi as high as $0.5$, provided that the decrease in $T_\alpha$ is compensated by $f_\alpha/f_c$ increasing to $1.2$ (see table \ref{table1}). For $\fesc=0.25,0.5$, within a $1\sigma$ error the simulated LFs can be reconciled with the observations for \avchi$\leq 0.5$, given that $f_\alpha/f_c$ increases to $1.6, 1.8$, respectively as shown in table 1; however, due to a decrease in the intrinsic \Lya luminosity and EW, the faint-end slope of the simulated LFs is always under-estimated with respect to the data for \avchi$\geq0.5$. As $\fesc$ increases to $0.75$ and above, the amplitude of the \Lya LFs can not be boosted up to the observed value, even for $f_\alpha=1$; although the \HII regions built by these galaxies are very large as a result of the large \HI ionizing photon escape fraction leading to large $T_\alpha$ values, the EW of the \Lya line prevents most of these galaxies from being classified as LAEs (see also Sec. \ref{homogenous_dust} above).

We therefore find a {\it three-dimensional degeneracy} between the IGM ionization state \avchi, the escape fraction of \HI ionizing photons ($\fesc$) and the clumped/inhomogeneous distribution of dust in the ISM of galaxies (governing $f_\alpha/f_c$) such that a decrease in $L_\alpha^{int}$ with increasing $\fesc$ can be compensated by a larger $T_\alpha$ and $f_\alpha/f_c$. However, we find that the \avchi values which reproduce the observations decrease with increasing $\fesc$ because an increasing $f_\alpha$ can not compensate for the decrease in the intrinsic \Lya luminosity. This degeneracy can be seen from Fig. \ref{fig_chi3D} where we show contours of $1-5\sigma$ deviations from the observations \citep{kashikawa2011}, obtained by computing the $\chi^2$ values for all the combinations of $\fesc =(0.05, .., 0.95)$, $\langle \chi_{HI} \rangle =(10^{-4}, .., 0.90)$ and $f_{\alpha}/f_c =(0.60,  .., 1.8)$, as shown in Table. \ref{table1}. 

\begin{table}
\begin{center}
\caption{For the \avchi value shown in column 1, for the $\fesc$ value shown as a subscript in each of the columns 2-4, we show the $f_\alpha/f_c$ ratio required to fit the simulated LAE \Lya and UV LFs to those observed by \citet{kashikawa2011} within a $1-\sigma$ error. Dashes indicate that the simulated LFs could not be matched to the observations within a $1-\sigma$ error, even for the maximum possible value of $f_\alpha=1$. }
\begin{tabular}{|c|c|c|c|}
\hline
\avchi&$(f_{\alpha}$/$f_c)_{0.05}$ &$(f_{\alpha}$/$f_c)_{0.25}$ &$(f_{\alpha}$/$f_c)_{0.5}$\\
\hline
0.50&1.2 &1.6 &1.8\\
0.25&0.8 &1.2  & 1.4\\
0.10&0.68 &1.0 & 1.4\\
0.01&0.68 &0.9 & 1.2\\
10$^{-4}$&0.60 &0.8 &1.2\\
\hline
\end{tabular}
\label{table1}
\label{table_bestfit}
\end{center}
\end{table}

The relation between \avchi and $\langle T_{\alpha} \rangle$ is complex due to the dependency upon the reionization topology as shown in \citet{dijkstra2011} \citep[see also Fig. \ref{fig_sources_XHI_transmission} and][] {jensen2013} and the assumed line shape \citep[e.g. Gaussian, double-peaked,][] {jensen2013,laursen2011,duval2014}.
For this reason we also show the three-dimensional degeneracy in terms of the mean transmission across all LAEs ($\langle T_{\alpha} \rangle_{LAE}$) for a given combination of $\fesc$ and $f_\alpha/f_c$ in Fig. \ref{fig_chi3D_transmission}. 

We note that for a given \avchi the mean transmission of all LAEs $\langle T_{\alpha} \rangle_{LAE}$ decreases for increasing $f_{\alpha}/f_c$: with rising $f_{\alpha}/f_c$ more galaxies, with lower values of $T_{\alpha}$, are identified as LAEs, leading to a decline in $\langle T_{\alpha} \rangle_{LAE}$. This effect can be seen for the maximum $\langle T_{\alpha} \rangle_{LAE}$ values, within a $4\sigma$ error, in each panel of Figure \ref{fig_chi3D_transmission} and it demonstrates the dependency of $\langle T_{\alpha} \rangle_{LAE}$ on $f_{\alpha}/f_c$.

From Figure \ref{fig_chi3D} and \ref{fig_chi3D_transmission}, we see that within a $1\sigma$ error, the observed \Lya and UV LFs can be reproduced for $\fesc$ values ranging between $0.05-0.5$ and \avchi$\simeq 10^{-4}-0.5$ or $\langle T_{\alpha} \rangle \simeq 0.38-0.5$, provided that the decreasing intrinsic \Lya luminosity (with increasing $\fesc$) and decreasing IGM transmission (with increasing \avchi) is compensated by an increasing $f_\alpha/f_c = 0.6$ to $1.8$. This implies that if dust is indeed clumped in the ISM of high-redshift galaxies such that the relative \Lya escape fraction can be as high as $1.8$ times the UV escape fraction, we can not differentiate between a universe which is either completely ionized or half neutral (or where $38-50$\% of LAE \Lya radiation is transmitted through the IGM) or an $\fesc$ ranging between 5-50\%. This parameter space is much larger than that allowed for a homogeneous dust distribution, where matching to the observations requires
$\fesc= 0.05$ and \avchi$\leq 0.1$ ($\langle T_{\alpha} \rangle_{LAE}\geq 0.43$).

% ***************************************************************************
\section{Conclusions} 
\label{sec6}
% ****************************************************************************
In this work we couple state of the art cosmological simulations run using GADGET-2 with a dust model and a radiative transfer code (pCRASH) to build a physical model of $z \simeq 6.7$ LAEs to simultaneously constrain the IGM reionization state, the escape fraction of \HI ionizing photons and ISM dust distribution (homogeneous/clumped).

We start by validating the properties of the simulated galaxy population against observations of high-redshift ($z \simeq 6-8$) LBGs. Identifying bound structures using the Amiga Halo Finder, simulated galaxies which are complete in the halo mass function ($M_h \geq 10^{9.2} M_\odot$), contain a minimum of $4N$ gas particles ($N = 40$) and a minimum of 10 star particles are identified as the ``resolved" population". For each such resolved galaxy, we use the mass, age and metallicity of each of its star particles to build the composite spectra. We use a dust model \citep{dayal2010a} that calculates the SNII dust-enrichment depending on the entire star formation history of a given galaxy, to obtain the associated escape fraction of UV photons. Once these calculations are carried out, galaxies with a dust-attenuated UV magnitude $M_{UV}\leq -17$ are identified as LBGs at $z \simeq 6-8$. Comparing the predictions by the simulation against LBG observations, we find that the slope and amplitude of the dust-attenuated UV LFs are in good agreement with the data at $z \simeq 6,7$ while the UV LF at $z \simeq 8$ requires no dust to match to the observations, such that $f_c \simeq (0.5, 0.6, 1.0)$ at $z \simeq (6,7,8)$, averaged over all LBGs. The theoretical Schechter parameters are found to be $\alpha = (-1.9\pm0.2, -2.0\pm0.2, -1.8\pm0.2)$ and $M_{UV,*} =(-19.8\pm0.3, -19.7\pm0.2, -19.9\pm0.3)$ at $z\simeq (6,7,8)$, and are consistent with those inferred observationally \citep{mclure2009, mclure2013,bouwens2011b}. We also find that the theoretical LBG stellar mass functions, SMDs and sSFRs are in excellent agreement with the data at $z \simeq 6-8$ (see Appendix \ref{a2}).

Once that the physical properties and dust enrichment of the simulated galaxy population have been validated against the mentioned observational data sets, we proceed to modelling LAEs. This requires further information on (a) the fraction of the \HI ionizing photons produced by a galaxy that are absorbed by ISM \HI ($1-\fesc$) and contribute to the \Lya emission line; the rest ($\fesc$) emerge out and contribute to reionization, (b) the relative effect of ISM dust on \Lya and UV photons ($f_\alpha/f_c$), and (c) the level to which the IGM is ionized by the galaxy population (\avchi), in order to calculate the IGM \Lya transmission. Since $\fesc$ remains poorly constrained with proposed values ranging between 1-100\% \citep[e.g.][]{ricotti-shull2000, gnedin2008, ferrara2013}, we use five different values of $\fesc = 0.05,0.25,0.5,0.75,0.95$ to run the radiative transfer code (pCRASH) in order to obtain various ``reionization" states of the simulated volume as it evolves from being fully neutral (\avchi$\simeq 1$) to completely ionized (\avchi$\simeq 10^{-4}$); pCRASH follows the time evolution of the ionization fractions (\HI and \HeII, \HeIII) and temperature given the spectra of the source galaxy population. Once these calculations are carried out, galaxies with $L_{\alpha}^{obs}\ge 10^{42}$erg s$^{-1}$ and $EW\ge20$\AA\ are identified as LAEs; the only free parameter of the model ($f_\alpha/f_c$) is then fixed by matching the simulated LFs to the observations. 

We start from the simplest case of considering the SNII produced dust to be {\it homogeneously distributed} in the ISM. The SNII extinction curve yields $f_\alpha/f_c=0.68$. Using this model, we find that for a given $\fesc$, the IGM \Lya transmission increases with decreasing \avchi, with this increase being most important for the smallest galaxies; as \avchi decreases, even small galaxies are able to ionize a large enough \HII region around themselves to be visible as LAEs. Hence, the \Lya LF becomes {\it steeper} with decreasing \avchi for any given $\fesc$ value. Further, for a given \avchi, although the ionization fields look very similar for different $\fesc$ values, the degree of ionization ($\chi_{HI}$) in any cell increases with $\fesc$ due to an increase in the \HI ionizing photon emissivity. This leads to a slight increase in $T_\alpha$ with $\fesc$ for a given \avchi. However, this increase is more than compensated by the decrease in the intrinsic \Lya luminosity with increasing $\fesc$ (as less photons are available for ionizations in the ISM), as a result of which the \Lya LF decreases slightly in amplitude with $\fesc$ for a given \avchi. Finally, comparing simulated \Lya and UV LFs with observations,
we find that for a homogeneous ISM dust distribution and $\fesc=0.05$, we can {\it constrain \avchi$\leq 0.1$ or $\langle T_{\alpha} \rangle_{LAE} \geq 0.43$}.

We also explore the scenario of dust being inhomogeneously distributed/clumped in the ISM of high-redshift galaxies, boosting up the ratio of $f_\alpha/f_c$. For a given \avchi value, an increase in $\fesc$ leads to a drop in the amplitude of the \Lya LF due to a decrease in $L_\alpha^{int}$, as mentioned above. In the clumped dust scenario, this decrease can be compensated by an increase in $f_\alpha$. For \avchi$\simeq 0.5$, $\fesc$ values as high as 0.5 can be reconciled with the data, given that $f_\alpha/f_c$ increases from $\simeq 0.6$ to $1.8$ as $\fesc$ increases from $0.05$ to 0.5. We thus find a {\it three-dimensional degeneracy} between \avchi, $\fesc$ and $f_\alpha/f_c$ such that a decrease in $L_\alpha^{int}$ with increasing $\fesc$ can be compensated by a larger $T_\alpha$ and $f_\alpha/f_c$. 
Due to the model dependency (e.g. Ly$\alpha$ line profile, reionization topology) of the \avchi-$\langle T_{\alpha} \rangle$ relation we also rephrase our result by converting the \avchi values to the corresponding mean transmission values for LAEs ($\langle T_{\alpha}\rangle_{LAE}$).
Within an $1\sigma$ error, we find that allowing for clumped dust, the data can be reproduced for a wide parameter space of $\langle T_{\alpha} \rangle_{LAE}\simeq 0.38-0.50$ or \avchi$\simeq 0.5-10^{-4}$, $\fesc \simeq 0.05-0.5$ and $f_\alpha/f_c\simeq0.6-1.8$, i.e., if dust is indeed clumped in the ISM of high-redshift galaxies such that the relative \Lya escape fraction can be as high as $1.8$ times the UV escape fraction, we can not differentiate between a universe which is either completely ionized or half neutral or where on average $38-50$\% of the Ly$\alpha$ radiation is transmitted, or has an $\fesc$ ranging between 5-50\%. We caution the reader that although the constraints in terms of $\langle T_{\alpha} \rangle_{LAE}$ account for the model dependent relation between \avchi and $\langle T_{\alpha} \rangle$, the value of $\langle T_{\alpha} \rangle_{LAE}$ sensitively depends on the galaxies that are classified as LAEs for a given combination of $\fesc$, \avchi and $f_\alpha/f_c$.

We now discuss some of the caveats involved in this work. Firstly, an enhancement of the escape fraction of Ly$\alpha$ photons with respect to continuum photons via the Neufeld effect \citep{neufeld1991} in a clumpy ISM remains controversial: on the one hand \citet{hansen-oh2006} showed that such enhancement was possible if the \Lya photons encountered a certain number of ``clumps". On the other hand, \citet{laursen2013} have shown that enhancing the ISM \Lya escape fraction requires very specific conditions (no bulk outflows, very high metallicity, very high density of the warm neutral medium and a low density and highly ionized medium) that are unlikely to exist in any real ISM. 

Secondly, as a consequence of simulating cosmological volumes we are unable to resolve the ISM of the individual galaxies to be able to model the Ly$\alpha$ line profile that emerges out of any galaxy. We have therefore assumed the emergent Ly$\alpha$ line to have a Gaussian profile with the width being set by the rotation velocity of the galaxy. However, several observational \citep{tapken2007,yamada2012} and theoretical works \citep{verhamme2008,jensen2013} have found the emerging Ly$\alpha$ line profile to be double-peaked; this profile results in higher IGM Ly$\alpha$ transmission values compared to a Gaussian profile  \citep{jensen2013}. An increase in $T_\alpha$ due to such a double-peaked profile allows our theoretical LFs to fit the observations with slightly lower $f_\alpha/f_c$ values.

Thirdly, we assume a constant escape fraction $\fesc$ for all galaxies. Although the value of $\fesc$ and its dependence on galaxy properties (such as the stellar mass) are only poorly known, recent work hints at an $\fesc$ that decreases with increasing mass \citep{ferrara2013, paar2011}. If $\fesc$ is indeed higher for low-mass galaxies, the ionization topology would become more homogeneous with the \HII regions being built by low-mass (high-mass) galaxies being larger (smaller) than the constant $\fesc$ case. This would have two effects: on the one hand the IGM Ly$\alpha$ transmission would be enhanced (decrease) for low-mass (high-mass) galaxies, leading to a steeper \Lya LF. On the other hand, due to its direct dependence on $\fesc$, the intrinsic Ly$\alpha$ luminosity would decrease (increase) for low-mass (high-mass) galaxies, leading to a flattening of the \Lya LF. The relative importance of these two effects is the subject of a work currently in preparation.

Fourthly, we assume that dust in high-redshift galaxies is solely produced by type II supernovae. However, as shown by \citet{valiante2009}, AGB stars can start contributing to significantly, and even dominate, the dust budget in regions of very high star formation (e.g. QSO/quasars) after $150-500$ Myrs. However, a significant part of this extra dust could be destroyed by forward SN shocks \citep{bianchi-schneider2007} that have also not been considered in this work. 

In ongoing calculations, our aim is to see whether the parameter space allowed by ``clumped" dust can be narrowed with a combination of LAE clustering and through 21cm brightness maps that will be directly testable with upcoming observations with the Subaru Hyper-Suprime Camera and LoFAR.

% ***************************************************************************
\section*{Acknowledgments} 
% ****************************************************************************
The authors thank the referee M. Dijkstra for his insightful and constructive comments, Gustavo Yepes for providing the hydrodynamical simulation and Benedetta Ciardi for a collaboration in developing  pCRASH.
This work has been performed under the HPC-EUROPA2 project (project number: 228398) with the support of the European Commission - Capacities Area - Research Infrastructures. 
AH thanks the European trans-national HPC access programme for a 8-week visit to the Institute of Astronomy (Edinburgh) where a part of this work was carried out. PD acknowledges the support of the European Research Council. 
The hydrodynamical simulation has been performed on the Juropa supercomputer of the J\"ulich Supercomputing Centre(JSC).

\bibliographystyle{mn2e}
\bibliography{vis3}

% **************************************************************************
 
\appendix

% ***************************************************************************
\section[]{Comparing the simulations with LBG observations} 
\label{a2}
% ****************************************************************************

% ************************************************************************************************
\subsection{Stellar mass functions and stellar mass density}
\label{smf}
% *************************************************************************************************

\begin{figure}
\begin{center}
\includegraphics[width=0.47\textwidth]{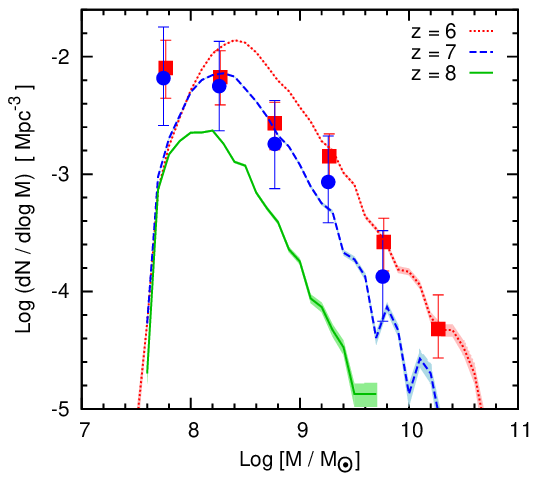}
\caption{Stellar mass function for LBGs with $M_{UV}\leq -18$ at $z\simeq 6-8$. The dotted (red), dashed (blue) and solid (green) lines represent the simulated stellar mass functions for $z \simeq 6, 7$ and 8, respectively. The shaded areas denote the respective Poisson errors. Squares (circles) represent the observational stellar mass functions at $z\simeq 6$ ($7$) inferred by \citet{gonzalez2011}, corrected for completeness but not for dust.}
\label{stelmassfn}
\end{center}
\end{figure}

\begin{figure}
  \center{\includegraphics[width=0.47\textwidth]{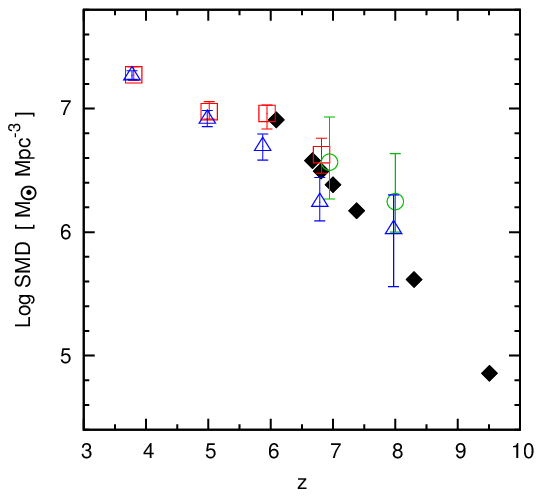}}
  \caption{Total stellar mass density for LBGs with $M_{UV}\leq- 18$. Filled black points represent the simulated SMDs; Poissonian errors are too small to be visible on the plot. The other symbols show the observed SMD values inferred by \citet[][empty red squares]{gonzalez2011}, \citet[][empty green circles]{labbe2010a}, \citet[][empty green circles]{labbe2010b}, \citet[][empty blue triangles]{stark2013}.}
  \label{smd}
\end{figure}

We build simulated LBG mass functions by summing up LBGs on the basis of their stellar mass, and dividing this by the width of the bin and the simulated volume; for consistency with observations, this calculation is carried out for all LBGs with $M_{UV}\leq -18$. As expected from the hierarchical model, small mass systems are the most numerous, with the number density decreasing with increasing mass. Further, the mass function shifts to progressively lower masses with increasing redshifts, as fewer massive systems have had the time to assemble. Both trends can be seen clearly from Fig. \ref{stelmassfn}: at $z \simeq 6$, galaxies with $M_* \simeq 10^{8.5} \Msun$ are two orders of magnitude more numerous as compared to more massive galaxies with $M_* \simeq 10^{10.5}\Msun$. Secondly, while there are systems as massive as $10^{10.7}\Msun$ in the simulated volume at $z \simeq 6$, such systems  had not the time to evolve by $z \simeq 8$ where the most massive systems have $M_* \simeq 10^{9.4}\Msun$. From the same figure, we see that the slope and amplitude of the simulated stellar mass functions are in agreement with the observations for $M_* \geq 10^{7.5}\Msun$; this number marks the limit of our simulation resolution for galaxies containing at least 4N gas particles and 10 star particles, as a result of which, the number density drops below it.  

As a result of the stellar mass functions shifting to progressively lower masses and number densities with increasing redshift (see Fig. \ref{stelmassfn}), the total SMD in the simulated volume decreases with increasing redshift, as shown in Fig. \ref{smd}, falling by a factor of about 40 from $10^7 \Msun \, {\rm Mpc^{-3}}$ at $z \simeq 6$ to $10^{5.4}\Msun \, {\rm Mpc^{-3}}$ at $z \simeq 9.5$. As expected from the agreement between the simulated stellar mass functions and those inferred observationally by \citet{gonzalez2011}, the simulated SMD values are also in agreement with the observations. However, as pointed out by \citet{schaerer2010} and \citet{stark2013}, including the effects of nebular emission can lead to a decrease in the observationally inferred stellar mass values; this effect causes a slight discrepancy between the SMD values obtained from our model, and those inferred by \citet{stark2013} at $z \simeq 6$, as shown in Fig. \ref{smd}.

\begin{figure}
  \center{\includegraphics[width=0.47\textwidth]{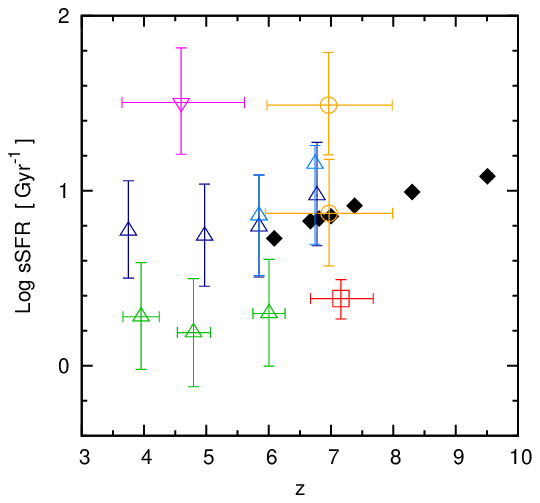}}
  \caption{Specific star formation rate for LBGs with $M_{UV}\leq - 18$. Filled black points represent the simulated sSFRs; Poissonian errors are too small to be visible on the plot. Observed sSFR values plotted have been taken from: \citet[][empty red squares]{gonzalez2010}, \citet[][empty green triangles]{stark2009}, \citet[][empty magenta triangles]{yabe2009}, \citet[][empty orange circles]{schaerer2010}, \citet[][empty blue triangles]{stark2013}. \label{SSFR}}
\end{figure}

% ************************************************************************************************
\subsection{Specific star formation rates}
\label{ssfr}
% *************************************************************************************************
The specific star formation rate (sSFR) is an excellent indicator of the current SFR compared to the entire past star formation history of the galaxy. This quantity also has the advantage that it is relatively independent of assumptions regarding the IMF, metallicity, age and dust content since both the SFR and stellar mass are affected similarly by these parameters. Recently, a number of observational groups have suggested that the sSFR settles to a value consistent with $2-3 \, {\rm Gyr^{-1}}$ at $z \simeq 3-8$ 
\citep{stark2009,gonzalez2010,mclure2011,labbe2012,stark2013}. This result is quite surprising given that theoretically, the sSFR is expected to trace the baryonic infall rate that scales as $(1+z)^{2.25}$ \citep{neistein2008}. However, \citet{labbe2012} and \citet{stark2013} have shown that at least a part of this discrepancy can be accounted for by the addition of nebular emission lines; indeed, the sSFR can rise by a factor of about 5 between $z \simeq 2$ and $z \simeq 7$ when nebular emission is taken into account.

Physically, the least massive galaxies tend to have the highest sSFR: as a result of their low $M_*$ values, even a small amount of SF can boost their sSFR compared to that of more massive systems \citep[see also][]{dayal2012, dayal2013}. Since the mass function progressively shifts to lower masses with increasing redshifts, and smaller mass galaxies have larger sSFR values, the sSFR rises with increasing redshift, from $sSFR \simeq 5 \, {\rm Gyr^{-1}}$ at $z \simeq 6$ to $sSFR \simeq 12 \, {\rm Gyr^{-1}}$ at $z \simeq 9.5$, in accordance with other theoretical results \citep{dayal2013,biffi2013,salvaterra2013}. We note that in the redshift range of overlap ($z \simeq 6-7$), the theoretically inferred sSFR values are in agreement with observations.  

\begin{figure*}
 \center{\includegraphics[width=1.0\textwidth]{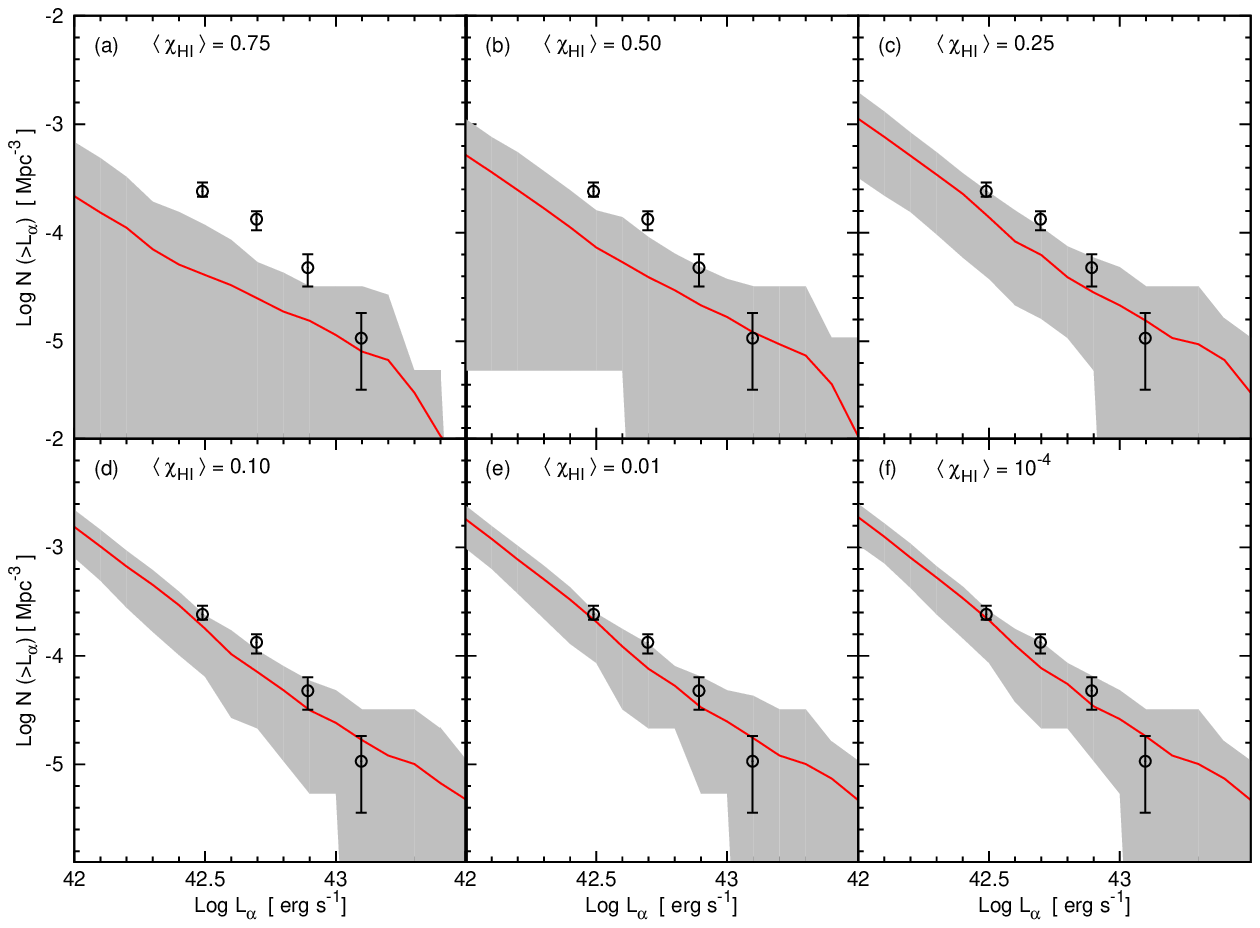}}
  \caption{Simulated cumulative \Lya LFs and the associated cosmic variance for homogeneous dust ($f_{\alpha}/f_c=0.68$) and $\fesc=0.05$. The panels show the results for different neutral hydrogen fractions, $\langle \chi_{HI} \rangle \simeq 0.90$, $0.75$, $0.50$, $0.25$, $0.10$, $0.01$, $10^{-4}$, as marked. In each panel, the solid red line shows the \Lya LF for the whole simulation box of volume $(80 h^{-1} {\rm Mpc})^3$ with the gray shaded areas showing the associated cosmic variance obtained by splitting the entire simulation box into 8 equal sub-volumes. In each panel, the open circles show the \Lya LFs at $z\simeq6.5$ observed by \citet{kashikawa2011}. As the IGM becomes more ionized, the value of $T_\alpha$ increases in all sub-volumes, leading to a decrease in the cosmic variance at the faint-end of the LF; the variance at the bright-end arises due to the low numbers of massive galaxies found in any given sub-volume.}
  \label{fig_variance}
\end{figure*}

% ***************************************************************************
\section[]{Cosmic variance} 
\label{a1}
% ****************************************************************************

In Fig. \ref{fig_variance} we show the cosmic variance associated with the simulated \Lya LFs. The cosmic variance is estimated by computing the \Lya LFs for $8$ sub-volumes of our simulation volume, such that each sub-volume is $1.87\times 10^5$ Mpc$^3$ and thus comparable to the volume observed by \citet{kashikawa2011}. As \avchi evolves from an IGM which is predominantly neutral at \avchi $\simeq 0.75$ to one wherein reionization is complete (\avchi$\simeq 10^{-4}$), the value of $T_\alpha$ increases in all sub-volumes, leading to a decrease in the cosmic variance at the faint-end of the LF; the variance at the bright-end arises due to few objects being massive/luminous enough to occupy these bins in any given sub-volume.

% **************************************************************************

\label{lastpage} 
\end{document}